\documentclass{aa}
\usepackage{graphicx}
\usepackage{txfonts}
\usepackage{longtable}
\usepackage{natbib}

\begin{document} 


\title{Galactic Bulge Microlensing Optical Depth from EROS-2\thanks{
Based on observations made with the MARLY
  telescope at the European Southern Observatory, La Silla, Chile.}}
\author{
C. Hamadache\inst{1},
L. Le Guillou\inst{1}\thanks{Now at Instituut voor Sterrenkunde,
    Celestijnenlaan 200 B,  B-3001 Leuven, Belgium},
P.~Tisserand\inst{1}\thanks{Now at 
Research School of Astronomy and Astrophysics,
Australian National University,
Mount Stromlo Obs.,
Cotter Rd., Weston, ACT 2611, Australia},
C.~Afonso\inst{1}\thanks{Now at Max-Planck-Institut f\"ur Astronomie,
Koenigstuhl 17,D-69117 Heidelberg, Germany
},
J.N.~Albert\inst{2},
J.~Andersen\inst{5},
R.~Ansari\inst{2},
\'E.~Aubourg\inst{1},
P.~Bareyre\inst{1},
J.P.~Beaulieu\inst{3},
X.~Charlot\inst{1},
C.~Coutures\inst{1,3},
R.~Ferlet\inst{3},
P. Fouqu\'e\inst{7,8},
J.F.~Glicenstein\inst{1},
B.~Goldman\inst{1}\thanks{Now at  Max-Planck-Institut f\"ur Astronomie,
Koenigstuhl 17,D-69117 Heidelberg, Germany},
A.~Gould\inst{6},
D.~Graff\inst{6}\thanks{Now at Division of Medical Imaging Physics,
Johns Hopkins University
Baltimore, MD 21287-0859, 
USA},
M.~Gros\inst{1},
J.~Haissinski\inst{2},
J.~de Kat\inst{1},
\'E.~Lesquoy\inst{1,3},
C.~Loup\inst{3},
C.~Magneville\inst{1},
J.B.~Marquette\inst{3},
\'E.~Maurice\inst{4},
A.~Maury\inst{8}\thanks{Now at San Pedro de Atacama Celestial Exploration,
Casilla 21, San Pedro de Atacama, Chile},
A.~Milsztajn \inst{1},
M.~Moniez\inst{2},
N.~Palanque-Delabrouille\inst{1},
O.~Perdereau\inst{2},
Y.R. Rahal\inst{2},
J.~Rich\inst{1},
M.~Spiro\inst{1},
A.~Vidal-Madjar\inst{3},
L.~Vigroux\inst{1,3},
S.~Zylberajch\inst{1}
\\   \indent   \indent
The EROS-2 collaboration
}    
\institute{
CEA, DSM, DAPNIA,
Centre d'\'Etudes de Saclay, 91191 Gif-sur-Yvette Cedex, France
\and
Laboratoire de l'Acc\'{e}l\'{e}rateur Lin\'{e}aire,
IN2P3 CNRS, Universit\'e de Paris-Sud, 91405 Orsay Cedex, France
\and
Institut d'Astrophysique de Paris,
UMR7095 CNRS, Universit\'e Pierre \& Marie Curie,
98~bis Boulevard Arago, 75014 Paris, France
\and
Observatoire de Marseille,
2 place Le Verrier, 13248 Marseille Cedex 04, France
\and
The Niels Bohr Institute, Copenhagen University, Juliane Maries Vej 30,
DK2100 Copenhagen, Denmark
\and
Department of Astronomy, Ohio State University, Columbus,
OH 43210, U.S.A.
\and
Observatoire Midi-Pyr\'en\'ees, Laboratoire d'Astrophysique (UMR 5572), 
14 av. E. Belin, 31400 Toulouse, France
\and
European Southern Observatory (ESO), Casilla 19001, Santiago 19, Chile
}
\date{Received;accepted}

\authorrunning{C. Hamadache et al.}
\titlerunning{Microlensing Optical Depth towards the Galactic
Bulge from EROS-2}

\def\lsim{{\lesssim}}
\def\au{{\rm AU}}
\def\etal{{et al.}}
\def\eros{{\sc eros}}
\def\macho{{\sc macho}}
\def\lmc{{\sc lmc}}
\def\smc{{\sc smc}}
\def\ie{{\em i.e.}}
\def\tempest%
{\begin{array}{ccc}
1 & 1 & 1 \\
1 & 1 & 1 \\
4 & 3 & 8
\end{array}}
\def\gsim{{{}_>\atop{}^{{}^\sim}}}
\def\lsim{{{}_<\atop{}^{{}^\sim}}}
\def\kms{{\rm km}\,{\rm s}^{-1}}
\def\kpc{{\rm kpc}}
\def\e{{\rm E}}
\def\rel{{\rm rel}}
\def\btheta{{\vec\theta}}
\def\bmu{{\vec\mu}}
\def\bpi{{\vec\pi}}
\def\Teff{{T_{\rm eff}}} 
\def\msun{\rm M_\odot} 
\def\deg{\rm deg}
\def\day{\rm d}
\def\te{t_{\rm E}}

\abstract
{
We present a new
EROS-2 measurement of the microlensing optical depth toward the
Galactic Bulge.
Light curves of $5.6\times 10^{6}$ clump-giant stars
distributed  over  $66\,\deg^2$ of the Bulge were
monitored during seven Bulge seasons.
120 events were found with apparent amplifications greater than 1.6 
and Einstein radius crossing times in the range
$5\,{\rm d}<t_\e <400\,{\rm d}$.
This is the 
largest existing sample of clump-giant events and the first to include
northern Galactic fields.
In the Galactic latitude range $1.4\degr<|b|<7.0\degr$,
we find
$\tau/10^{-6}=(1.62\,\pm 0.23)\exp[-a(|b|-3 \,{\rm deg})]$
with $a=(0.43\,\pm0.16)\deg^{-1}$.
These results are in good agreement with our previous measurement,
with recent measurements of  the MACHO and  OGLE-II
groups, and  with predictions of Bulge models.

}

\maketitle           

\section{Introduction} 
\label{section:introduction}

Gravitational microlensing of stars is an important tool for 
constraining the quantity and characteristics of 
faint compact objects between the stars and the observer.
In microlensing events, the lensing object passes near the line of sight
towards the background star, causing a 
transient magnification of the star's primary image
as well as creating a secondary image.
At Galactic scales, neither the image separation nor the
image size are normally resolvable, so the only easily observable
effect during a microlensing event 
is an apparent transient amplification of the star's flux.
The characteristic timescale giving the effective duration
of the amplification
is proportional to the square root of the lens mass and
inversely proportional to the transverse velocity.
For Galactic stellar lenses, the timescales are on the order of 
a few weeks.

Microlensing searches were originally proposed
\citep{Pac1986}
as a tool for detecting dark matter in galactic halos.
Searches for the lensing of stars in the Magellanic Clouds 
by the MACHO \citep{MACHOMC} and EROS-2 \citep{EROSMC,tisserand}
projects have 
placed constraints on the  fraction of the Milky Way halo that can
be comprised of faint compact objects.
Searches for objects in the halo of M31 are also being performed
with candidate events reported by the
VATT-Columbia \citep{Ugle04},
WeCAPP \citep{Riff03},
AGAPE  \citep{PH2005}, 
MEGA \citep{mega2005} and
Nainital \citep{nainital} collaborations.
The AGAPE and MEGA collaborations presented efficiency calculations
allowing them to constrain the content of the  M31 and Milky Way halos.

Microlensing has also been used as a tool to investigate
compact objects in the visible regions of the Milky Way.
Besides  the Galactic Bulge that is the subject
of this paper, microlensing of stars in the spiral arms of the Galaxy
has also been studied \citep{DER01}.

Microlensing surveys toward the Galactic Bulge were first
proposed \citep{PAC91,GRI91} as a probe of ordinary
stars in the Galactic Disk, though
it was soon realized \citep{KIRA94} that lensing by stars in the Bulge
itself is of comparable importance.
The optical depth, i.e. the probability that 
at a given time a  star at a distance $D_{\rm s}$
 is amplified by more than a factor
1.34 is 
\begin{equation}
\tau \;=\; \frac{4\pi G D_{\rm s}^2}{c^2}\int_0^1 \rho(x)x(1-x)dx \;,
\label{optdepthint}
\end{equation}
where $\rho$ is the mass density of lenses and $x=D_{\rm l}/D_{\rm s}$
is the ratio between the lens and source distances.
Very qualitatively, the integral for the Disk contribution 
to the optical depth is
\begin{equation}
\tau_{\rm disk} \;\sim\; \frac{GM_{\rm disk}}{c^2 h_{\rm disk}} \;,
\label{optdepthdisk}
\end{equation}
where $M_{\rm disk}$ and $h_{\rm disk}$ are the total mass
and scale height of the Disk lensing population. Lenses in the
Bulge are near the source star  so
the Bulge contribution is
\begin{equation}
\tau_{\rm bulge} \;\sim\; \frac{GM_{\rm bulge}}{c^2 R_{\rm bulge}} 
            \;\sim\; \frac{G\rho_{\rm bulge}R^2_{\rm bulge}}{c^2 } \;,
\label{optdepthbulge}
\end{equation}
where $M_{\rm bulge}$, $\rho_{\rm bulge}$ and   
$R_{\rm bulge}$ are the
mass, density and line of sight thickness of the Bulge 
lensing population. 
For Bulge stars near the direction of 
the Baade window ($\ell=1\degr ,b=-3.9\degr$),
the most recent calculations 
\citep{EVA02,Bissantz,HanGould03,woodmao} 
give total optical depths
in the range $1<\tau/10^{-6}<2$,
with about 60\% of the rate being due to lensing by Bulge stars and
the remainder by  Disk stars.
The expected optical depth naturally has a strong  dependence
on the Galactic latitude, $b$, falling
typically from $\tau\sim5\,\times10^{-6}$ at $(l\sim0,|b|=1\degr) $ to
$\tau\sim5\,\times10^{-7}$ at $(l\sim0,|b|=6\degr )$.

The first
measurements of the optical depth towards the Galac\-tic
Bulge by the OGLE
collaboration 
\citep{UDA94b, UDA00, WOZ01}, the MACHO collaboration 
\citep{ALC97, ALC00} and
the MOA collaboration \citep{ sumi2003}
yielded optical depths significantly higher
than these estimates, suggesting that a fundamental revision
of Galactic models might be  necessary \citep{BIN00}. 
However, 
the interpretation of the first results
was difficult because of  two effects.  
First, stars in the direction
of the Galactic Bulge are not necessarily located in the Bulge itself so
foreground and background contamination must be taken into account.
Second, and more importantly,
photometry of the numerous faint stars in the crowded
Galactic Bulge fields is complicated by ``blending'' where the
reconstructed stellar flux receives contributions from more
than one star.  
This makes
the effective number of monitored stars greater than the
number of cataloged stars and generates a complicated
relationship between real and reconstructed amplifications.

These problems can be largely avoided by considering
only microlensing of clump giant stars.  Such stars
are identified by their well-defined position in the color-magnitude
diagram. 
This position ensures that they are most likely in the Galactic
Bulge.  Their large flux also makes blending problems relatively
unimportant.  

The first published optical depth using clump giants was
based on only 13 events
observed by the MACHO group \citep{ALC97}.  
The optical depth,
 $3.9^{+1.8}_{-1.2}\times 10^{-6}$ at 
$\langle l,b\rangle=(2.55\degr ,-3.64\degr )$, was
still considerably higher than expectations.
Since then, measurements using clump giants have tended to give
results in better agreement with models.
Based on 16 events,
the EROS-2 collaboration \citep{Afonso} gave an optical depth
of $0.94\pm0.29\times 10^{-6}$ at 
$\langle l,b\rangle =(2.5\degr ,-4.0\degr)$.  
The MACHO group \citep{machocgpop} recently presented their results
using 62 events with clump giant sources.  Their optical depth is
$\tau=2.17^{+0.47}_{-0.38}\times\,10^{-6}$ at
$\langle l,b\rangle =(1.5\degr,-2.68\degr)$, in good agreement
with calculations and with their earlier preliminary
results \citep{POP00}.
The OGLE-II group \citep{sumiogle}
recently used 33 events to 
obtain  $\tau=2.55^{+0.57}_{-0.46}\times 10^{-6}$
at $\langle l,b\rangle =(1.16\degr,-2.75\degr)$, 
again in agreement with models.

        In this paper we present a measurement of the
optical depth for microlensing toward the Galactic Bulge
based on the totality of the EROS-2 clump-giant data. 
More details on this analysis can be found in \citet{hamadache}.
The 120 events used comprise the largest sample of clump-giant
events studied to date.
Section \ref{section:data} describes the data collection and
processing required to produce the light curves.
Section \ref{clumpsec} describes the definition of the
clump-giant sample to be used to measure the optical depth.
Section \ref{section:event_selection}  lists the criteria used
to isolate a sample of microlensing candidates.
Section \ref{section:efficiencies} describes the calculation of
the detection efficiency, the optical depth and the timescale distribution.
Section \ref{section:effect_blending} discusses the possibility
that stellar blending may affect the results. 
Finally, Section \ref{conclusionsec} compares our results with those
of other groups and with the predictions of Galactic models.

\section{Data}
\label{section:data}

The data were acquired by  EROS-2  with the 
$1\,{\rm m}$ MARLY telescope
at La Silla, Chile. The imaging was done simultaneously by two cameras,
using a dichroic beam-splitter \citep{BAU97}. 
Each camera was composed of a mosaic of 
eight 2048$\times$2048 pixel LORAL CCDs, with a 
pixel size of $0.\hskip-2pt''6$ yielding a
field of $0.7\degr (\alpha)\times
1.4\degr(\delta)$.  
Each camera observed through one of two non-standard, broadband
filters, $R_{\rm eros}$ and $B_{\rm eros}$. The filter
$R_{\rm eros}$ covers the range 620-920 nm and has  
a mean wavelength near that of Cousins
$I$,  while $B_{\rm eros}$ covers the range 420-720 nm 
and has a mean wavelength between
Johnson/Cousins $V$ and $R$. 
The photometric calibration was obtained directly for
20\% of our fields  by matching our star catalogues with those of the
OGLE-II collaboration
\citep{ogleIIphotometry}.  
To a precision  of $\sim 0.1\,{\rm mag}$, we find
\begin{equation}
R_{\rm eros} \;=\; I_{\rm ogle}\;, \hspace*{5mm}
B_{\rm eros} \;=\; V_{\rm ogle}-0.4(V-I)_{\rm ogle} \;. 
\end{equation}
The calibration is sufficiently uniform that it can be extended
with confidence to the remaining fields.  We note, however, that
none of our conclusions depend on the calibration precision.

\begin{figure}
 \centering
 \includegraphics[width=7.8cm]{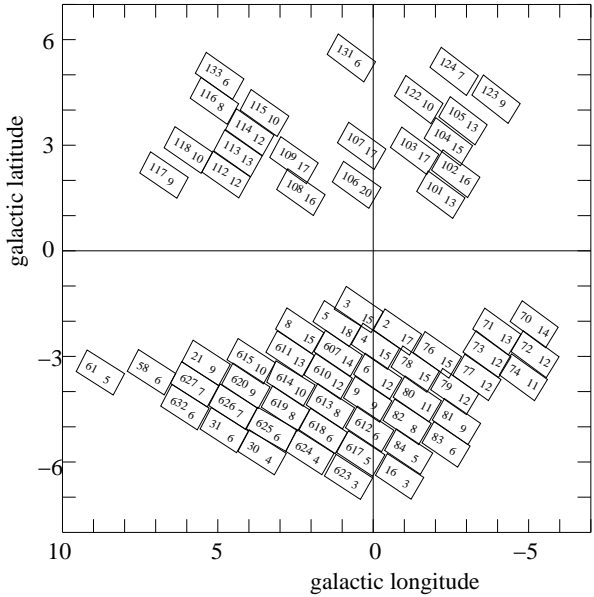}
 \caption{Map of the EROS-2 Bulge fields in Galactic coordinates. 
A total of 66
fields were monitored.  The first number in each field is the field
number and the second is the number of clump giants in the field divided
by $10^4$.
The field 121 with $4\times\,10^4$ clump giants is 
outside the plot at $(l,b)=(11.5,2.5)$.
}
\label{fig:eros2_bulge_fields}
\end{figure}

\begin{figure}
 \centering
 \includegraphics[width=7.8cm]{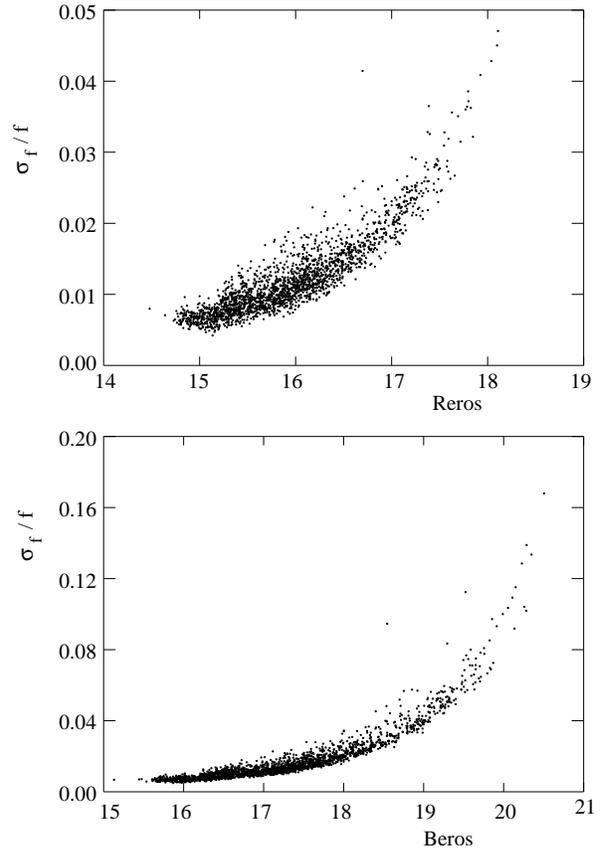}
 \caption{
The dispersion, $\sigma/f_{\rm base}$,
of clump-giant light curves about the base flux
as a function of $R_{\rm eros}$ and of $B_{\rm eros}$. 
Each point represents the median dispersion of clump giants
for one subfield consisting of a CCD-quadrant.
}
  \label{baselineresfig}
\end{figure}

The observations reported here concern 66 fields, 
monitored between July 1996 and October 2002.
Figure \ref{fig:eros2_bulge_fields} shows the location of the 66 fields in
Galactic coordinates, $(l,b)$.  Twenty-two of the fields are in
the north ($b>0$) and 44 in the south.
The corresponding data set contains $6\times 10^7$ pairs of 
light curves, of which
$5.6\times 10^6$ are Bulge clump giants that we use in the present analysis. 

The fields are observable at La Silla from mid-February to mid-October.
An average of one point every four nights was taken for each field
during the observing season though the sampling frequency 
occasionally reached more than one point per night 
at mid-season when the 
Galactic Bulge is observable throughout the  night.

The image photometry was performed with soft\-ware spe\-cific\-ally 
designed for
crowded fields, PEIDA (Photo\-m\'etrie et \'Etude d'Images Destin\'ees
\`a l'Astrophysique) \citep{ANS96}.
The first step 
consisted of creating  reference images for each 
of the 2112 subfields
of size  $\sim 0.035\,\deg^2$, i.e. one quarter of a CCD.
A reference image was constructed by
adding 15 high-quality images of the subfield.
Star catalogs for each subfield were established using
the reference image.
Photometry on individual images was then performed by
imposing the stellar positions found on the reference
image.

A first estimate of the  
uncertainty of individual flux measurements is
that due to the photometric fit taking into account
only photon counting statistics.
This clearly underestimates the uncertainty since it ignores
other contributions important in crowded field photometry, e.g.
imperfect knowledge of the point-spread-function (PSF).
The factor by which the uncertainties must be increased
is determined by comparing individual flux measurements
with the flux measurements of the same star on the reference image.
The fit errors are   
increased by a magnitude and image dependent (but star independent)
factor,
chosen so that for each image
the  distribution of
the difference between flux and
reference flux is consistent with that expected from
the renormalized flux uncertainties.
By construction, this procedure yields, for most stars,  
a $\chi^2$ per degree of freedom near unity for a fit
to a time-independent flux.

The uncertainties on flux measurements
are reflected in the dispersions of light curves
about the baseline flux  (estimated from the mode of the flux).  
Figure \ref{baselineresfig} shows,
for each subfield, the median dispersion  for clump giants in the
two colors.  The photometric precision is typically 2\% but degrades
considerably for $B_{\rm eros}$ in highly absorbed fields.
We will be concerned only with microlensing apparent 
amplification factors $>1.6$
so the precision in most fields is sufficiently good to yield  high
quality light curves in both colors.
The 20\% of the microlensing 
candidates with the lowest quality light curves were
also photometered with our differential photometry package \citep{triton}
so as to confirm their microlensing origin.

Examples of light curves containing
microlensing events are
shown in Figs. \ref{cdlfig1}-\ref{cdlfig4}.

\section{Clump giant selection}
\label{clumpsec}

\begin{figure}
 \centering
 \includegraphics[width=7.8cm]{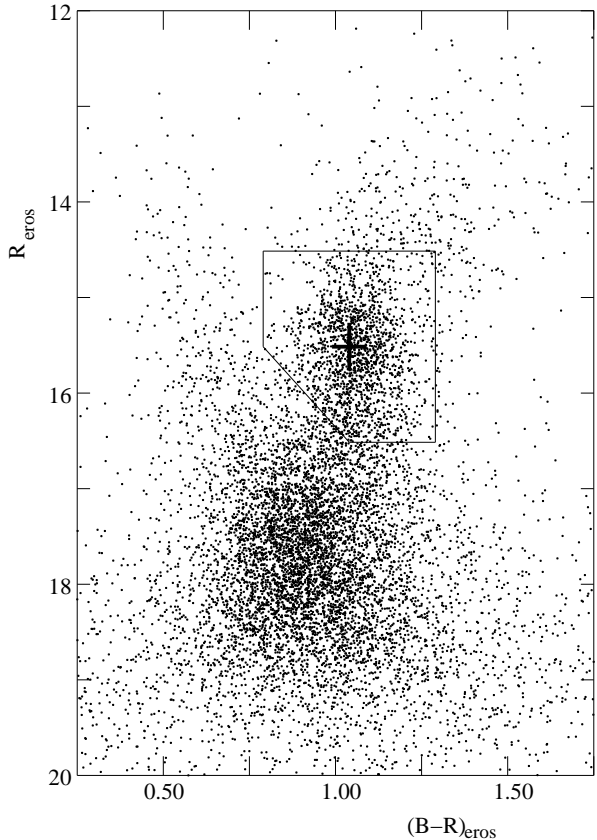}
 \caption{
A typical color-magnitude diagram of 
a subfield
(field 610, CCD 4 quadrant 2).
The cross shows the fitted center of the clump and the thin
lines show the cuts defining the clump giants used in this analysis.
}
  \label{cmdfig}
\end{figure}

The sample of clump-giants to be studied was selected 
using the color-magnitude diagrams for each CCD-quadrant
subfield.  Figure \ref{cmdfig} shows a diagram of
a typical subfield.  The center
of the clump was determined by modeling the density of stars
in  color-magnitude space 
as a power law representing
a smooth background plus a Gaussian representing the clump.
The Gaussian had the form
\begin{displaymath}
{\rm density} \;\propto\; 
\exp\left[-\frac{((B-R)_{\rm eros} - \alpha)^2}{2\sigma_{\alpha}^2}\right]  
\end{displaymath}
\begin{displaymath}\hspace*{10mm}\times
\exp\left[-\frac{(R_{\rm eros}+1.9(B-R)_{\rm eros} - \beta)^2}
{2\sigma_{\beta}^2}\right]  \;
\end{displaymath}
where 
$\alpha$ ($\sigma_{\alpha}$) and $\beta$ ($\sigma_{\beta}$) 
are the desired clump centers (widths) in the space
defined by $(B-R)_{\rm eros}$ 
and the reddening-free magnitude
$R_{\rm eros}-1.9(B-R)_{\rm eros}$.
The dereddened magnitude is used
so as to be relatively insensitive to 
variable reddening across the subfields.

\begin{figure}
 \centering
 \includegraphics[width=7.8cm]{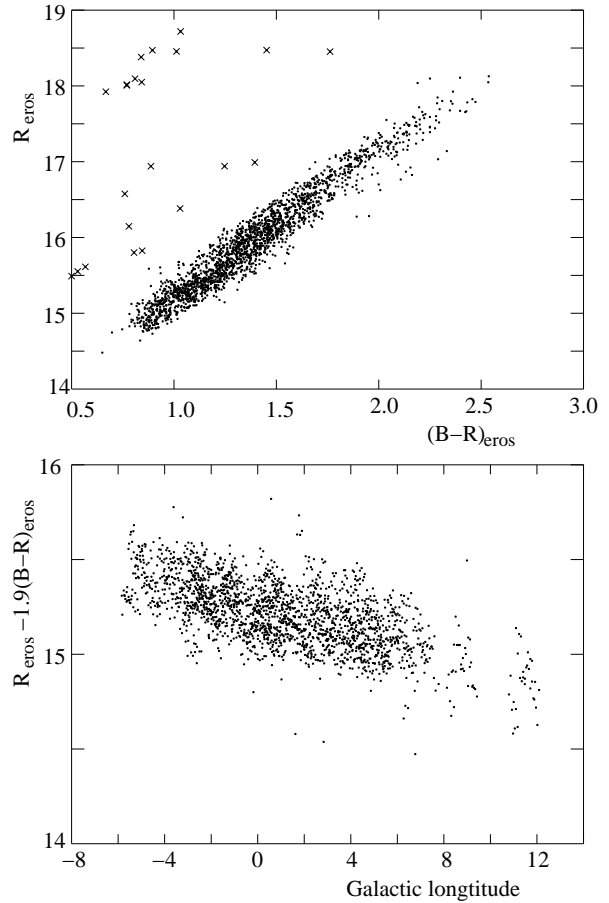}
 \caption{The top panel shows
$R_{\rm eros}$ of the clump center vs $B_{\rm eros}-R_{\rm eros}$ of the clump 
for the 2109 subfields (CCD quadrants). 
(Three quadrants were not reconstructed.)  
The distribution follows the expected
reddening law.  Quadrants marked with a $\times$ were eliminated from
further consideration.
The bottom panel shows the dereddened magnitude as a function of
Galactic longitude.  The increasing flux
with longitude is believed to be 
due to the bar structure since stars with $l<0$
are farther than stars with $l>0$.
}
  \label{clumpposfig}
\end{figure}

\begin{figure}
 \centering
 \includegraphics[width=7.8cm]{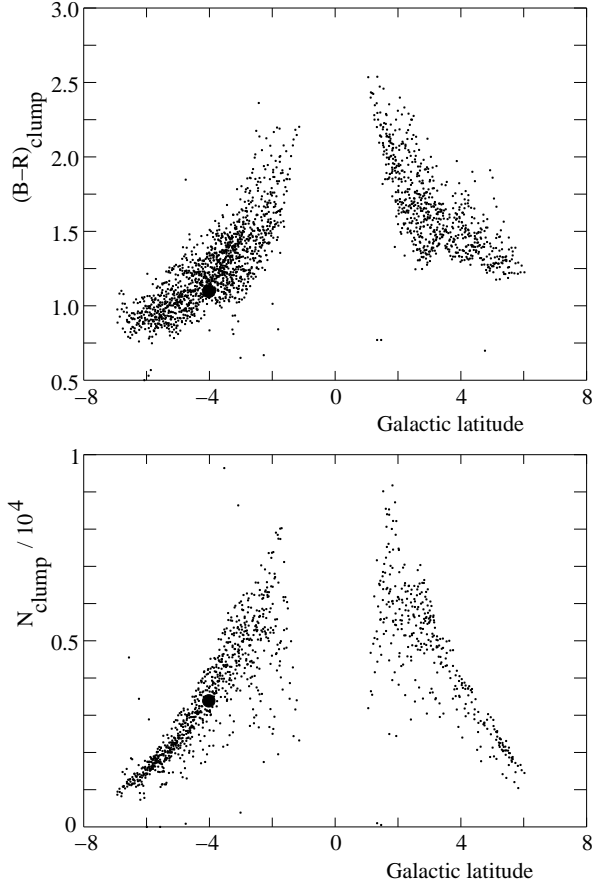}
 \caption{
The 
mean clump color (top panel) 
and the
number of clump giants per CCD-quadrant (bottom panel) 
as a function of Galactic
latitude for the longitude range $-3\degr<\ell <3\degr$.
The number of giants is symmetric about $b=0$ while northern
fields are more reddened than southern fields.
The large dot marks the position of the subfield containing
the Baade Window.}
  \label{clumpnumfig}
\end{figure}

Figure \ref{clumpposfig}a shows the fitted $R_{\rm eros}$
of the clump center as a function of the fitted $(B-R)_{\rm eros}$
of the clump center.
The distribution follows the expected reddening law indicating
that variations of clump position are due to differential reddening.
The 46 subfields that are outliers in this plot are excluded from
further consideration leaving 2063 subfields.
All 2063 color-magnitude diagrams were visually inspected
to ensure that reasonable values for the clump position were found.
We note that, in spite of the large variation of clump position,
the clump is
visually distinct in each of the 2063 diagrams.
The variations of clump position lead to varying photometric
precisions (Figure \ref{baselineresfig}) but this is taken into
account in the microlensing detection efficiency (Section \ref{section:efficiencies}).

Figure \ref{clumpposfig}b shows the dereddened clump magnitude
as a function of Galactic longitude, $\ell$.  
The longitude dependence is believed to be caused by the
bar structure of the Bulge where
stars with $l>0$ 
are brighter than stars with $l<0$ because they are nearer
\citep{oglebar}.
The 0.38 mag difference over $12\degr$ in longitude corresponds
to a bar orientation of $49\degr\pm8\degr$ with
respect to the line of sight to the Galactic Center.
This angle is in agreement with the original
OGLE-I results \citep{oglebar} and
with the recent infrared star counts from GLIMPSE \citep{glimpse}.

Figure \ref{clumpnumfig}a shows the clump color as a function of
Galactic latitude.  The northern fields are more absorbed and reddened
than the southern fields.

The cuts used to define our sample of clump giants, illustrated
in Figure \ref{cmdfig}, are
\begin{equation}
|R-R_{\rm clump}|\,<\,1\;, 
\end{equation}  
\begin{equation}
|(B-R)-(B-R)_{\rm clump}|\,<\,0.25 \;,
\end{equation}  
\begin{equation}
R-R_{\rm clump}\,>\, -4.[(B-R)-(B-R)_{\rm clump}]\,-1\;,
\end{equation} 
where $R_{\rm clump}$ and $(B-R)_{\rm clump}$ define the
clump center. 
The number of clump giants as a function of
Galactic latitude is shown in Figure \ref{clumpnumfig}b.
As expected, the number falls rapidly with increasing distance
from the Galactic equator.  In spite of the difference in 
absorption between northern and southern fields, there is
no obvious asymmetry between counts of northern and southern
clump giants, as expected.

\section{Event Selection}
\label{section:event_selection}

After the production of the
light curves, microlensing candidates 
among the $5.6\times10^6$ clump giants were found by
the  procedures described in this section.  
After elimination of
images with instrumental problems or poor seeing, the first step was
to calculate the mode of the red and blue fluxes that was
used as a first estimate of the baseline fluxes.

The light curves were then subjected to a ``filter'' that selects
light curves having one or more groups of points with fluxes
sufficiently above 
the baseline flux, $f_{\rm base}$.
Such groups were initiated with any point $i$ having a flux $f_i$
greater than $f_{\rm base}+2.5\sigma_i$ where $\sigma_i$ is the
estimated uncertainty of $f_i$.
The groups  ended with the  first
subsequent sub-group of three points all with 
$f_i<f_{\rm base}+2\sigma_i$.
The relative significance of the groups on a light curve
was defined by
\begin{displaymath}
LP_N \;=\; N\log2 \;-\; \sum_{i=1}^N \log 
\left[ {\rm erfc}
\left( \frac{f_i-f_{\rm base}}{\sqrt{2}\sigma_i} \right)
\right] \;,
\end{displaymath}
where $N$ is the number of points in the group.
For each red or blue light curve, groups having at least 5 points
with $f_i>f_{\rm base}+2.5\sigma_i$
were then ordered by
decreasing significance.
A light curve was then retained for further analysis if 
one of the two most significant red groups had
a  temporal overlap of at least 20\% with
one of the two most significant blue groups.

Roughly 3\% of the light curves pass this filter.
The filter's effective threshold is sufficiently low that, for clump
stars, the filter passes essentially all microlensing events
with apparent amplifications greater than 30\%
occurring during well-sampled parts of the observing period.
Significant losses of efficiency occur due to bad weather, equipment
failure and limited visibility of the Galactic Bulge at the beginning
and end of each season.

Light curves passing the filter were then fit in each color with the
microlensing light curve corresponding to  
uniform motion of an unblended  point source and single lens.
The total flux as a function of time is
\begin{equation} 
f(t) \; = \; f_{\rm base} \frac{u^{2}+2}{u\sqrt{u^{2}+4}}
\label{noblendeq}
\end{equation}
where the impact parameter $u(t)$ is
\begin{equation}
u^{2}(t)  =  u_{0}^{2} + \frac{(t-t_0)^2}{t_\e^2} \; .
\label{u2defeq}
\end{equation} 
For each color, there are  4 fitted  parameters:
the baseline flux $f_{\rm base}$,
the time  of maximum apparent
amplification $t_0$, the
impact parameter at maximum amplification
normalized to the Einstein ring radius $u_0=u(t_0)$, 
and finally, the microlensing event duration, i.e. 
the Einstein ring radius crossing time $t_\e=r_\e/v_{\rm t}$ for a transverse
relative velocity $v_{\rm t}$. 
The last two quantities depend on the Einstein radius,
$r_\e^2=4GMD_{\rm l}(D_{\rm s}-D_{\rm l})/c^2D_{\rm s}$, 
where $M$ is the
mass of the lens, 
and $D_{\rm l}$ and $D_{\rm s}$ are the distances to the lens and source,
respectively.

While all transient flux variations can be fitted  more or less
successfully with the above expression, ideal microlensing events are 
characterized by achromaticity, i.e. the same $u_0$, $t_\e$ and
$t_0$ in all pass-bands.  Because of the
rarity of the phenomenon, one generally also expects the
events to be not repeated for a given star.
A final important characteristic is that the detection efficiency
corrected distribution of $u_0$ is flat.
These characteristics guide the
final selection of microlensing candidates.  This selection is
based on seven criteria, C1-C7, 
applied to the parameters of the microlensing fit performed
on light curves passing the filter.  
The cuts we use
lead to a detection efficiency that is relatively
independent of $u_0$ so
the flat distribution of $u_0$ is only moderately  distorted.  
On the other hand, the cuts necessarily result in an
efficiency that  depends on $t_\e$ since very short events
have a significant chance of falling between observations.
The cuts are also sufficiently loose to
maintain a good efficiency for finding non-conventional
microlensing events, e.g. those due to binary lenses.

The first five criteria,
C1-C5, are applied independently to the parameters of the
fits for both bands and events are required to pass the criteria
in both.  
C1 requires that the time of maximum apparent amplification
be within the observing season so that the microlensing fit will
yield reliable lensing parameters:
\begin{displaymath}
{\rm C1:} \hspace*{10mm} t_0\;\in{\rm observing\;   season}
\end{displaymath}
The beginning and end of the seven observing seasons were
defined separately for each of the 66 fields 
by the first and last successful images for that field.
For each field,
only seasons with a mean sampling interval of less than 8.5 days
are used. 
Sub-fields covered by  CCD 2 were ignored for the first 4 seasons when the 
red CCD-2
was mostly out of order.

The second cut eliminates very long candidate events:
\begin{displaymath}
{\rm C2:} \hspace*{10mm}  t_\e \;<\; 400\,{\rm d}\;.
\end{displaymath}
A visual scan of events eliminated by this cut indicates that they
are generally low amplitude events  ($u_0>0.8$).
No high amplitude 
microlensing events ($u_0<0.5$) were seen.
Many of the $t_\e>400\,{\rm d}$ events
appear to be  simple baseline shifts,  
with most occurring 
when the telescope optics were realigned
at $t=JD-2,450,000=983$.  Most stars with such shifts
are near very bright stars 
and it is probable that the shifts are due to slight changes
in the wings of the PSF.

The third cut eliminates events due to known
equipment problems.
Most importantly, we eliminate
long events occurring before
$t=JD-2,450,000=900$ by requiring
\begin{displaymath}
{\rm C3:} \hspace*{10mm} t_e < 100\,{\rm d}\;{\rm if}\;
t_0<900\;.
\end{displaymath}
A visual scan of the $\sim5000$ events  eliminated with this cut indicates
that almost all are due to the aforementioned baseline shifts at  
$t=983$ which yielded events with $t_0\sim800$.
We also eliminated 48 events 
on CCDs 6 and 7 between days 2323 and 2343
that were due to an electronics problem.

The next cut requires sufficient sampling during the event
\begin{displaymath}
{\rm C4:} \hspace*{10mm} {\rm \geq 4\;points\; with}\; 
|t-t_0|<t_\e
\;,
\end{displaymath}
\begin{displaymath}
\hspace*{5mm} {\rm and} \hspace*{5mm} {\rm \geq 8\;points\; with}\; 
|t-t_0|<2t_\e
\;.
\end{displaymath}
This cut has the largest influence on  the $t_\e$ dependence of 
the efficiency.

Cut 5 is a very loose cut on the $\chi^2$ of the microlensing
fit calculated using only the points far from the time of maximum
amplification, 
$|t-t_0|>2t_\e$:
\begin{displaymath}
{\rm C5:} \hspace*{10mm} \chi_{\rm base}^2/N_{\rm dof} < 10\;.
\end{displaymath}
This cut
eliminates most periodic and chaotic variable stars but retains
single excursion events, even those  that fit badly the simple  light curve
(\ref{noblendeq}),
e.g. events due  to binary lenses.

Cuts 6 and 7 require reasonable agreement between the blue and
red light curves:
\begin{displaymath}
{\rm C6:} \hspace*{10mm} \frac{t_0({\rm blue})- t_0({\rm red})}
{\langle t_\e\rangle} < 0.4\;,
\end{displaymath}

\begin{displaymath}
{\rm C7:} \hspace*{10mm} \frac{t_\e({\rm blue})- t_\e({\rm red})}
{\langle t_\e\rangle} 
< 0.3\;.
\end{displaymath}
For the most part, these two cuts eliminate short time-scale excursions due
to photometric problems on single exposures.

\begin{figure}
 \centering
 \includegraphics[width=7.8cm]{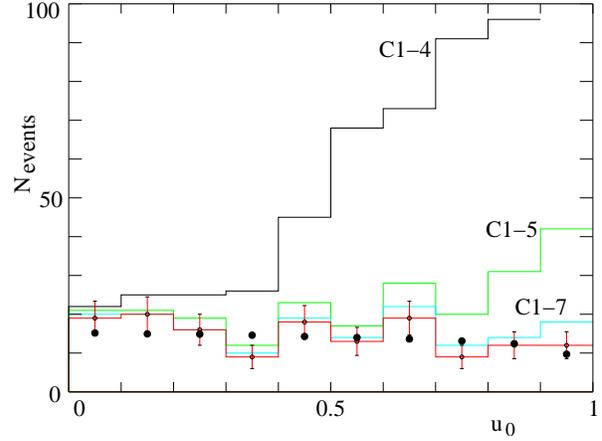}
 \caption{
The 
evolution of the $u_0$ distribution with the applied 
candidate selection criteria.  The first histogram is for
the 577 events with $u_0<1.0$ satisfying criteria C1, C2,C3  and  C4.  The
second is for the 234 events that satisfy the five criteria C1-C5  
and the third for 
the 165 events satisfying the seven criteria C1-C7. 
In the final histogram, 18 events have been eliminated whose
light curves show additional variations outside the main variation
or are clearly baseline shifts.
The dots give the Monte-Carlo distribution for microlensing events
satisfying all criteria.  
}
  \label{u0distfig}
\end{figure}

\begin{table}
\begin{center} \begin{tabular}{l l l  } \hline\hline
event type       & number & $\sum(t_\e/\epsilon)/{\rm total}$\\
\hline
\\
simple & 104 & 0.77\\
strong blends (B) &  5 & 0.02   \\
strong parallax (P)& 4 & 0.11   \\
binary non caustic (X)  & 2 & 0.02    \\
caustic (C) &  5 & 0.06 \\
single excursion (?) &  3 & 0.01\\
chaotic/periodic variable \\
or baseline shift & 6 & -\\
\\
\end{tabular}
\caption{
The tentative classifications of the 129 events satisfying cuts C1-C7 and
$u_0<0.75$.  The symbols in parentheses (B,P,X,C,?) show how these
classifications are designated in Figure \ref{c2fig} and 
Table \ref{eventtable1}.
The last column gives the relative weight of each classification if
they were used in the calculation of the optical depth using the $t_\e$
from the  simple-lens fit. (In reality, only 120 events are used, eliminating
the 3 single excursion (?) events and 6 chaotic/periodic variables
or baseline shifts.)
}
\label{scantable}
\end{center}
\end{table}

To understand the effect of each cut it is interesting to follow the
evolution of the $u_0$ distribution as successive cuts are applied.
The first distribution in Figure \ref{u0distfig} is for the 577 events
with $u_0<1.0$
that pass cuts C1, C2, C3 and C4.  The distribution is dominated
by events with $u_0>0.5$  corresponding to low amplitude variations.
This is indicative of contamination by low amplitude variable stars.
However, already with just these cuts, the distribution 
is flat for $u_0<0.4$ indicating that there is little background
for high apparent amplification microlensing events.

The second distribution is for the 234 events that also satisfy
C5.  This eliminates most of the variable stars, and the $u_0$ distribution
is now quite flat for $u_0<0.8$.  Application of cuts C6 and C7 eliminate
a few additional events, mostly near $u_0\sim1$, leaving  165
events with a distribution
that is flat for $u_0<1.0$.  

The 165 light curves were then visually  examined.
Most events fit well the simple microlensing light curve though
a significant fraction show interesting effects.
A summary of the scan conclusions is
given in Table \ref{scantable} for the 129 events with $u_0<0.75$ where
there is generally little ambiguity about the nature of the events.
Non-simple microlensing events include four  strong parallax 
events, i.e. those with light curves strongly affected by
the circular motion of the Earth (P).  There are  
five binary lens events with visible caustic
crossings (C).
Five events show strong source blending (B) and are discussed in more
detail in Section \ref{section:effect_blending}.
Two events (X) show achromatic deviations from the
simple lens curve that might be due either  to non-uniform motion
of the source in a binary system  (i.e. a so-called xallarap event) 
or to a binary lens without caustic crossing.
Six events  show chaotic or periodic variations
outside the primary variation and are therefore clearly
not microlensing events.  Finally
three events marked (?) show single excursions that fit badly the simple
lens curve but could conceivably be due to binary lenses.

\begin{figure}
 \includegraphics[height=.9\textheight]{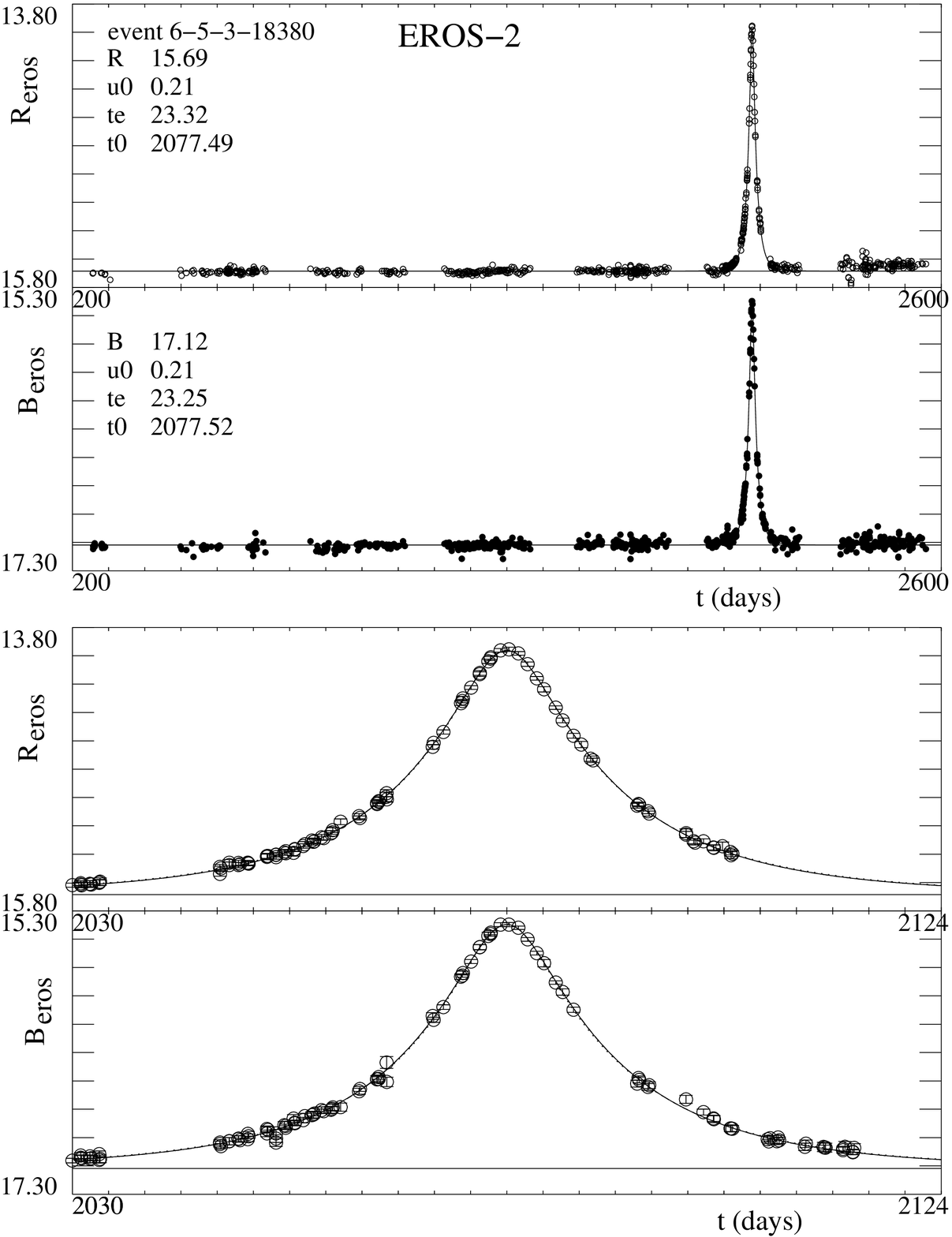}
 \caption{
The light curves of EROS-2 microlensing candidate 
6-5-3-18380.
Of the 120 candidates, this one has  the largest value of 
$\chi^2({\rm flat})-\chi^2({\rm microlensing})=221600$. 
The two top panels show $R_{\rm eros}$ and $B_{\rm eros}$
as a function of time (JD-2,450,000).  The two bottom panels show
a zoom for the time interval $t_0-2t_\e <t< t_0+2t_\e$.
}
  \label{cdlfig1}
\end{figure}

\begin{figure}
 \includegraphics[height=.9\textheight]{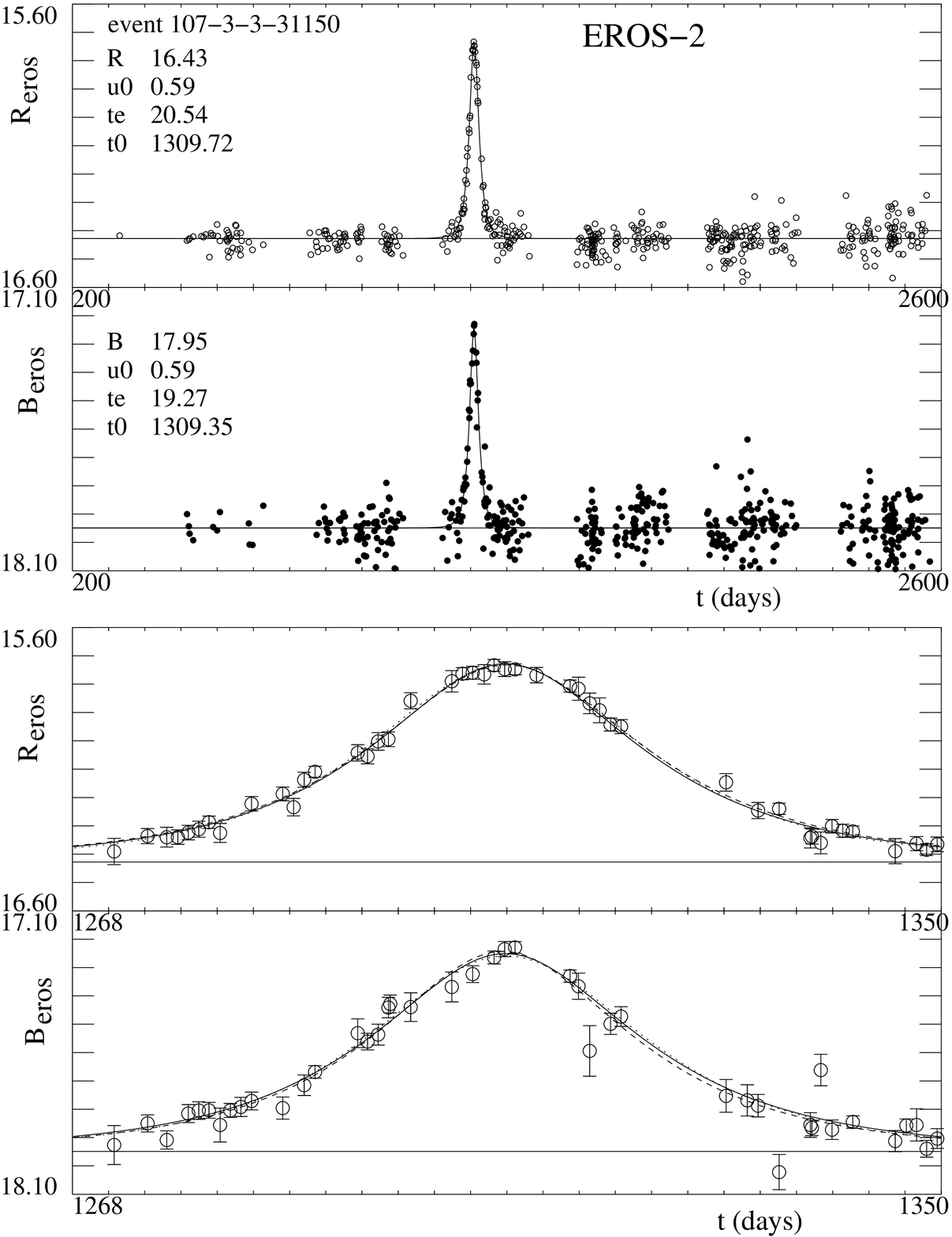}
 \caption{
The light curves of the EROS~2 microlensing candidate
107-3-3-31150. 
Of the 120 candidates, this one has the median value of 
$\chi^2({\rm flat})-\chi^2({\rm microlensing})=6810$.
The two top panels show $R_{\rm eros}$ and $B_{\rm eros}$
as a function of time (JD-2,450,000).  The two bottom panels show
a zoom for the time interval $t_0-2t_\e <t< t_0+2t_\e$.
On the bottom two panels, the solid line is the blend-free 
simultaneous fit for both
colors, the dashed line the blend-free single-color fit, 
and the dotted line the
simultaneous fit with blending.
}
  \label{cdlfigmedian}
\end{figure}

\begin{figure}
 \includegraphics[height=.9\textheight]{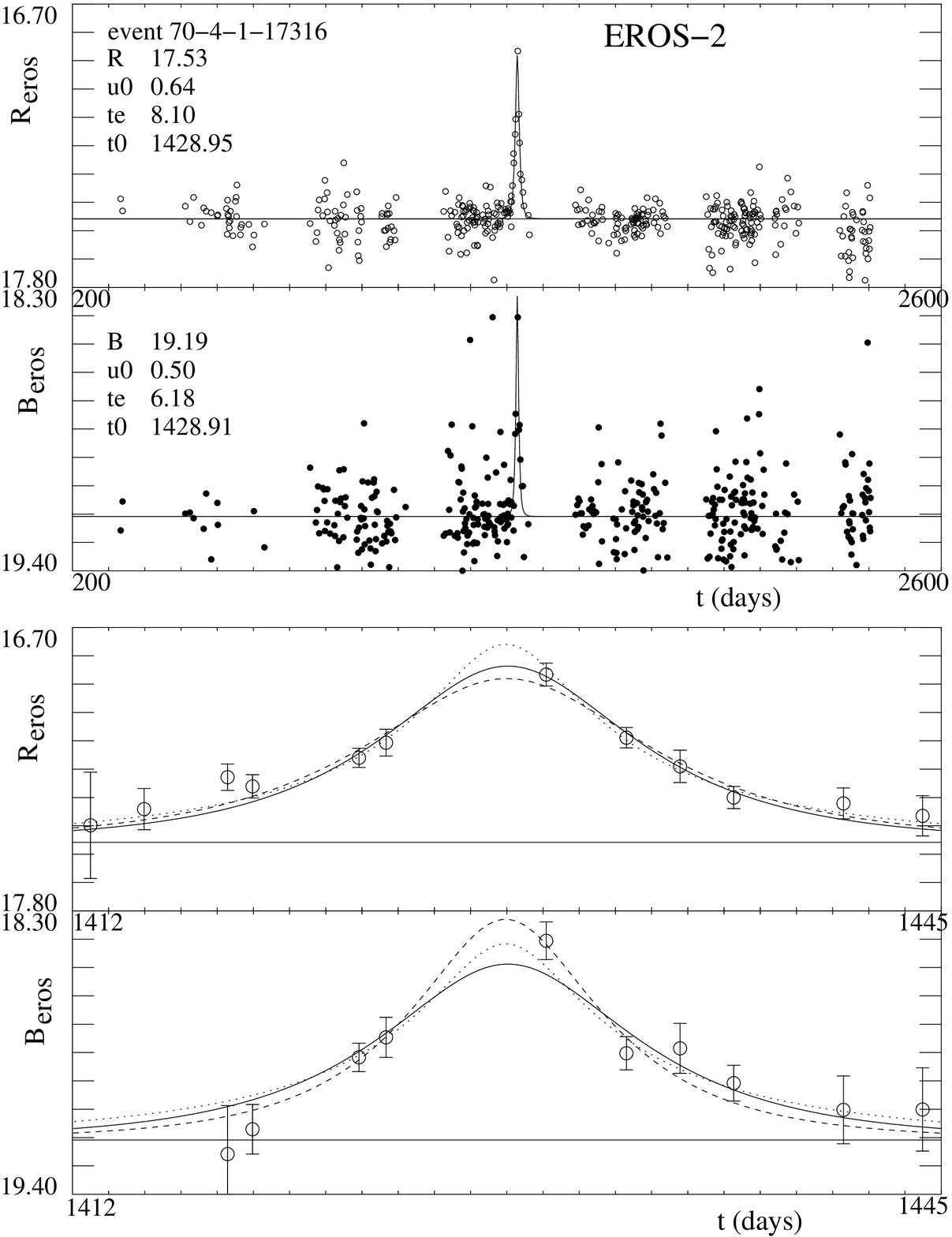}
 \caption{The light curves of the EROS~2 microlensing candidate
70-4-1-14316. 
Of the 120 candidates, this one has  the smallest value of 
$\chi^2({\rm flat})-\chi^2({\rm microlensing})=399$.
The two top panels show $R_{\rm eros}$ and $B_{\rm eros}$
as a function of time (JD-2,450,000).  The two bottom panels show
a zoom for the time interval $t_0-2t_\e <t< t_0+2t_\e$.
On the bottom two panels, the solid line is the blend-free 
simultaneous fit for both
colors, the dashed line the blend-free single-color fit, 
and the dotted line the
simultaneous fit with blending.
}
  \label{cdlfig4}
\end{figure}

The events with $u_0>0.75$ are visually not very
striking, and a non-negligible number were eliminated by scanning.
We therefore choose to eliminate these events so that 
for the remainder of this paper, we  consider only the 120 events
selected by C1-C7 and classified as microlensing events 
with $u_0<0.75$, i.e. maximum amplifications $A_{\rm max}>1.6$. 
The characteristics of the 120 events are listed in Table \ref{eventtable1}.
Figures \ref{cdlfig1}, \ref{cdlfigmedian} and \ref{cdlfig4}  show 
the light curves of three candidate events, the first with 
the most significant improvement in $\chi^2$ over the constant-flux
light curve, the second with the median improvement, and the third  with the
least significant improvement.
All of the 120 light curves can be found on our web page,
{\it http://eros.in2p3.fr}.

\begin{figure}
 \centering
 \includegraphics[width=7.8cm]{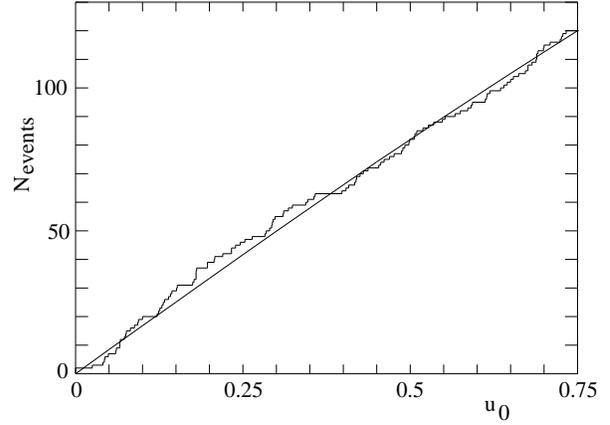}
 \caption{The cumulative $u_0$ distribution for the 120
selected events.  The smooth curve shows the efficiency-corrected 
flat $u_0$ distribution.
}
  \label{kolgofig}
\end{figure}

\begin{figure}
 \centering
 \includegraphics[width=7.8cm]{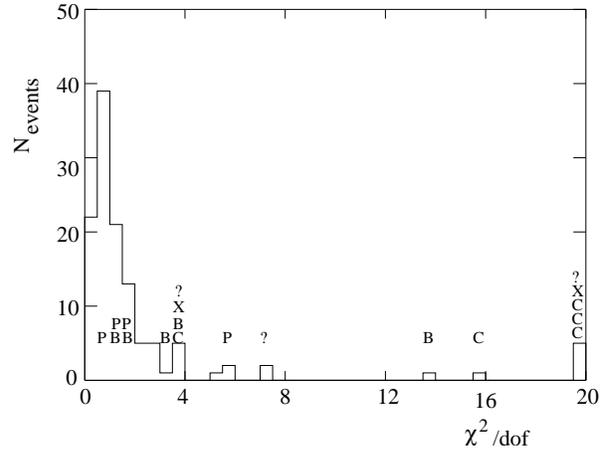}
 \caption{
The distribution of the reduced $\chi^2$ calculated
using points within the range $|t-t_0)|<2t_\e$.  
The 120 accepted events plus the 3 (?) events appear
in the plot.
The letters show the positions of the 19 non-simple
events as described in Table \ref{scantable}.
}
  \label{c2fig}
\end{figure}

\begin{figure}
 \centering
 \includegraphics[width=7.8cm]{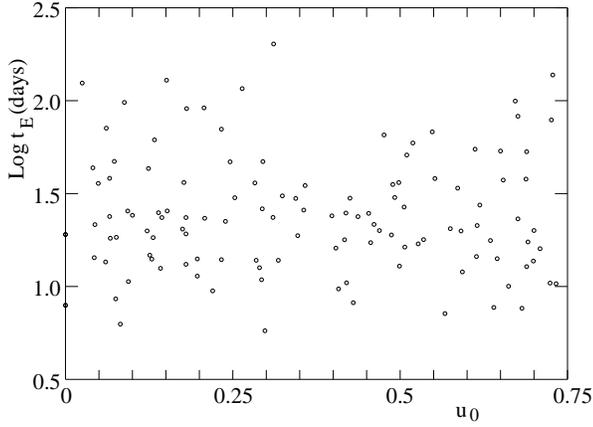}
 \caption{
The distribution of $(u_0,t_E)$ for the 120 events passing
all selection criteria with $u_0<0.75$.
}
  \label{u0tefig}
\end{figure}

\begin{figure}
 \centering
 \includegraphics[width=7.8cm]{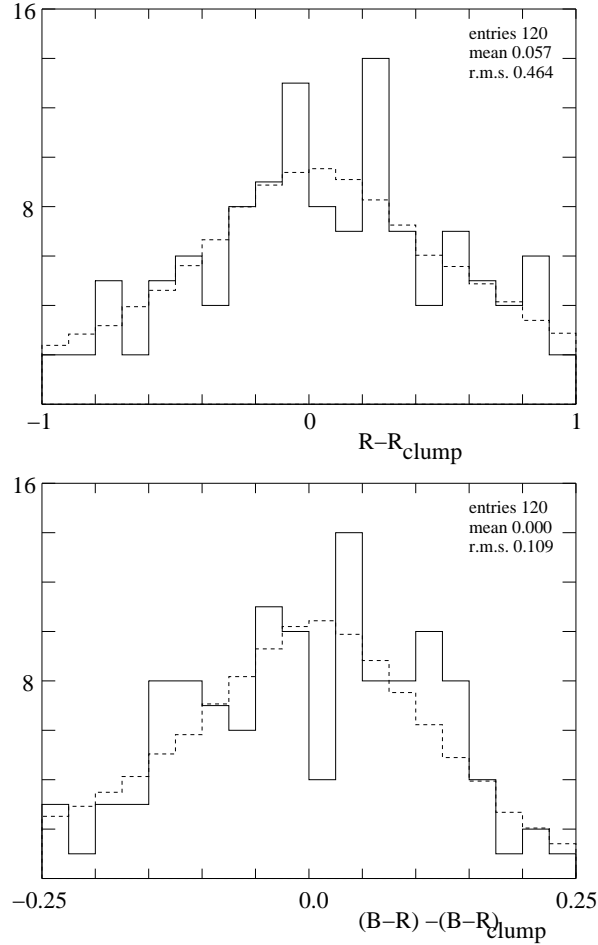}
 \caption{
The top panel shows 
the distribution of $R_{\rm eros}$ relative to the center
of the clump, $R_{\rm clump}$.
The bottom panel shows 
the distribution of $(B-R)_{\rm eros}$ relative to the center
of the clump, $(B-R)_{\rm clump}$.
In both cases, the dashed histogram shows the distribution of
random clump stars.
}
  \label{clumpfig}
\end{figure}

The $u_0$ distribution for the selected events  
is quite flat, as can be seen in Figure \ref{u0distfig}.
It agrees with the expected distribution from the 
Monte-Carlo simulation.  The fact that 
the observed and expected distributions
fall slightly near $u_0=1$ is due to the effect of the event filter that
requires a sufficient number of points significantly above the baseline.
Figure \ref{kolgofig} shows the cumulative $u_0$ distribution for
the 120 events and the
efficiency-corrected curve for 
a flat $u_0$ distribution (with the observed distribution of $t_\e$).
The Kolgomorov-Smirnov test gives a CL of 83\%.
As discussed in Section \ref{section:effect_blending}, the sample
may have a $\sim10\%$ contamination
at large $u_0$ by blended events due to the microlensing of faint
background stars.  Eliminating any randomly selected 10 events in the 
expected range $0.5<u_0<0.75$ yields a CL of 34\%, still quite
acceptable.

The distribution of $\chi^2$ per degree of freedom calculated
for points in the range $|t-t_0|<2t_\e$ is shown in Figure \ref{c2fig}.
The simple events have a reasonable distribution with a mean value
of 1.24, a quite satisfactory value considering the difficulty
of estimating photometric uncertainties. 
The other categories of events have, naturally, larger values of 
reduced $\chi^2$, especially caustic events (C).

Figure \ref{u0tefig} shows the scatter plot of $\log(t_\e)\, vs.\, u_0$.
The $u_0$ distribution is, as expected, not strongly  dependent on
$t_\e$.  Figure \ref{clumpfig} shows the distribution of event 
magnitude and color relative to the center of the clump.  The
events show a distribution that is very similar to that of the
ensemble of clump stars.

\begin{figure}
 \centering
 \includegraphics[width=7.8cm]{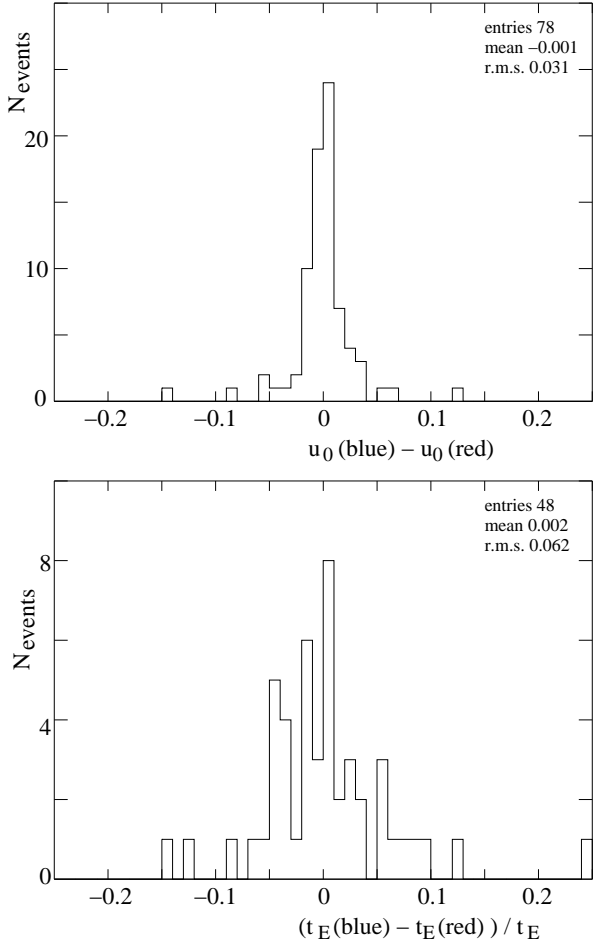}
 \caption{Tests for achromaticity of the events.  
The top
panel shows the difference between red and blue minimum impact
parameters, $u_0$,  for the 78 events with at least three points
in the range $t_0-t_\e/2 < t< t_0+t_\e/2$ for both red and
blue curves, ensuring a precise measurement of the maximum
amplification in both colors.
The bottom
panel shows the difference between red and blue $t_\e$
for the 48 events with at least three points
in the range 
$t_0-3t_\e/2 < t< t_0-t_\e/2$ and three points in the range
$t_0+t_\e/2 < t< t_0+3t_\e/2$ 
for both red and
blue curves, ensuring a precise measurement of $t_\e$
in both colors.
}
\label{chromfig}
\end{figure}

\begin{figure}
 \centering
 \includegraphics[width=7.8cm]{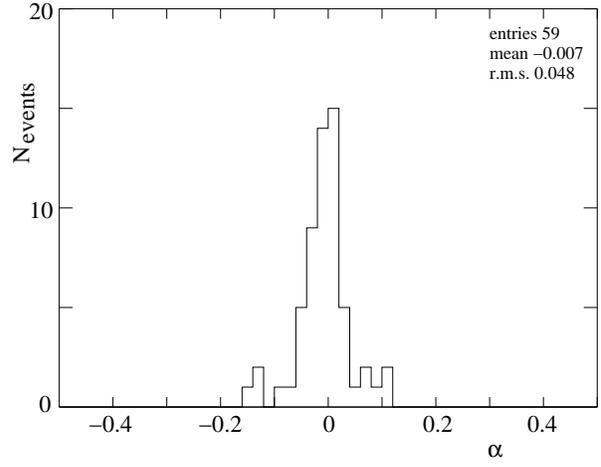}
 \caption{Test for time symmetry of the events.
The histogram is of the parameter $\alpha$ defined by (\ref{alphadefeq})
for events with $t_0$ at least 50 days away from both the beginning and the end
of the observing season, ensuring a good measurement of $\alpha$.
}
  \label{updnfig}
\end{figure}

The events in the sample are generally well described by
the light curve (\ref{noblendeq}), as indicated by  
the reduced $\chi^2$ distribution (Figure \ref{c2fig}).
Further indication that the events are indeed microlensing events
comes from their achromaticity and time symmetry.
Figure \ref{chromfig} shows that the events have  equal
$u_0$ and $t_\e$ in the blue and red as expected for microlensing
events with negligible blending.
Figure \ref{updnfig} shows that the events have equal rise 
and fall times.
This test for time symmetry was performed by 
fitting  the light curves with
the form (\ref{noblendeq})  with $t_\e$ in
(\ref{u2defeq}) modified by
\begin{equation}
t_\e \hspace*{2mm}\rightarrow \hspace*{2mm}
\Delta t \;=\; t_\e
\left[
1 \,+\, \alpha\, 
\arctan \left( \frac{t-t_0}{t_\e} \right)
\right] \;.
\label{alphadefeq}
\end{equation}
The parameter $\alpha$ describes the asymmetry of the light curve
with $\alpha=0$ corresponding to a symmetric (microlensing) curve
and $\alpha>0$  ($\alpha<0$) corresponding to rise times less than
(greater than) the fall time.  As an example, supernova light curves generally
give $\alpha> 0.3$.
The observed distribution (Figure \ref{updnfig})  is symmetrically
peaked at $\alpha=0$ as expected for microlensing events.

\section{Optical depth and $t_\e$ distribution}
\label{section:efficiencies}

\begin{figure}
 \centering
 \includegraphics[width=7.8cm]{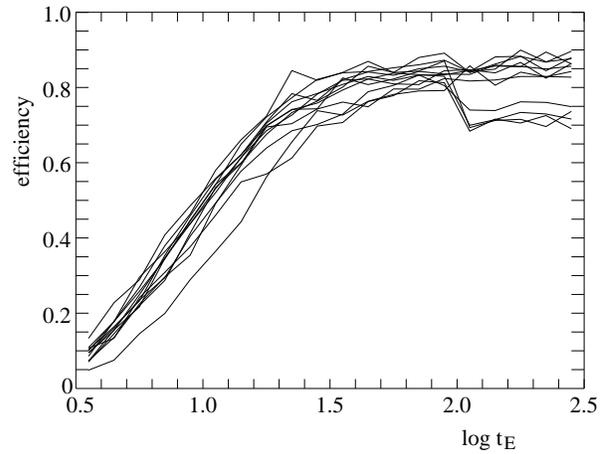}
 \caption{Detection efficiency for individual fields
as a function of the event time scale $t_\e$
(in days) for $u_0({\rm max})=0.75$.
For the sake of clarity, only 12 of the 66 fields are shown.
The discontinuity at $t_\e = 100\day$ is due to criterion C3
for fields that were observed during the first two seasons.
}
  \label{fig:eff_cg2_moy_1}
\end{figure}

To determine the optical depth 
and the distribution of $t_\e$, we must first  
evaluate the detection efficiency as a function of time scale $t_\e$ by
using Monte-Carlo simulated light curves. We superimpose
artificial microlensing events, with randomly generated parameters (impact
parameter, date of maximum amplification and time scale), on 
a random sample of real light curves.
These light curves are then subjected to the same microlensing
search as the original light curves and 
we determine the fraction that are 
recovered by our detection algorithm. 
More precisely, the detection efficiency as a function of $t_\e$
is given by the ratio between two numbers:

\begin{itemize}

\item  The number of events passing all selection criteria and with
the reconstructed $t_\e$ within the considered $t_\e$ bin and
with the reconstructed $u_0<0.75$.

\item  The number of generated microlensing events with
generated $t_\e$ within the considered $t_\e$ bin,
with generated $t_0$ within the observing period for the
field in question and
with the generated $u_0<0.75$.

\end{itemize}
Figure~\ref{fig:eff_cg2_moy_1} shows the efficiency 
as a function of $t_\e$ for a subsample of individual fields.
The efficiency is field-independent
at the 10\% level.
For $t_\e >100\,\day$ criterion C3 causes a slight discontinuity in 
the efficiency 
for fields that were well observed during the first two seasons.

   The microlensing optical depth is defined as the probability
that a given star, at a given time $t$, is magnified by at least 1.34,
i.e. with an impact parameter $u(t)<1$. The optical depth is then given
by   
\begin{equation}
\tau = \frac{\pi}{2u_0({\rm max})} 
\frac{
\sum_{i=1}^{N_{\rm ev}} t_{\rm E,i}/\epsilon(t_{\rm E,i})
}{
\sum_{j=1}^{N_{\star}} T_j
}\;, 
\label{eqn:theo_opt}
\end{equation}
where $N_\star$ is the number of monitored stars, $T_j$ is the
observation period for star $j$, 
$t_{\rm E,i}$ is the measured Einstein
crossing time of the $i$th candidate and $\epsilon(t_{\rm E,i})$ is the
detection efficiency.
For each event, the efficiency is taken from the corresponding
subfield consisting of one CCD (Figure \ref{fig:eff_cg2_moy_1}).

\begin{figure}
 \centering
 \includegraphics[width=7.8cm]{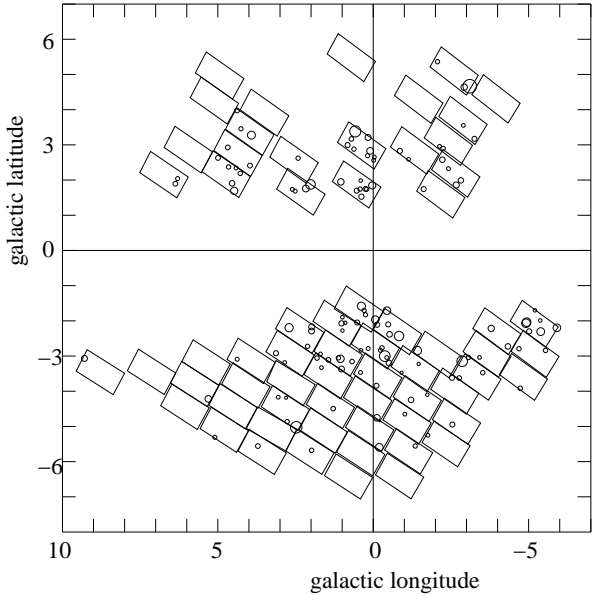}
 \caption{The location of the 120 candidates used for the calculation of the
   optical depth. The area of each circle is proportional to the
   contribution of each candidate to the optical depth.}
  \label{fig:cont_candidates}
\end{figure}

The denominator of (\ref{eqn:theo_opt}) is 
$6.62\times10^9{\rm star\cdot days}$.
Using the 120 events in Table \ref{eventtable1}, we find a
calculated optical depth averaged over all fields which is  
\begin{equation}
\tau = 1.68\,\pm 0.23 \times 10^{-6}\;, \hspace*{5mm} \langle |b| \rangle
\;=\; 3.43\degr \; .
\label{eqn:obs_opt}
\end{equation}
The uncertainty is mostly statistical, estimated following the prescription of
\citet{HAN95b} 
\begin{equation}
\sigma(\tau) = \tau\frac{\sqrt{<t_\e^2/\epsilon^2>}}{<t_\e/\epsilon>}
\frac{1}{\sqrt{N_{\rm ev}}}=
0.21\times 10^{-6}\; .
\end{equation}
To the statistical error we have added in quadrature a 5\%
systematic uncertainty due to blending effects, as discussed in 
Section \ref{section:effect_blending}.
The location and 
contribution of each of the 120 candidates to the measured optical
depth is shown in Figure~\ref{fig:cont_candidates}.

\begin{table*}
\begin{center} \begin{tabular}{l l r r l l l l  } \hline\hline
 range       & $\langle |b| \rangle$ & $N_{\rm stars}$  
& $N_{\rm ev}$ &
$\langle t_\e \rangle $ &  $\sigma(t_\e)$ & $\langle t_\e\rangle_{\rm cor}$ & $\tau/10^{-6}$ \\ 
\hline
\\
$   1.40<|b | <   7.00$ &    3.34 &  5569216 &  120 &    32.95 &    31.07 &  $    28.31\pm    2.84 $  &  $    1.68\pm    0.22 $ \\
\\
$   1.40<|b | <   2.00$ &    1.75 &   630884 &   25 &    29.17 &    21.85 &  $    24.88\pm    4.37 $  &  $    3.52\pm    1.00 $ \\
$   2.00<|b | <   2.50$ &    2.26 &   829123 &   22 &    39.29 &    28.20 &  $    30.91\pm    6.01 $  &  $    2.38\pm    0.72 $ \\
$   2.50<|b | <   3.00$ &    2.76 &   976707 &   24 &    22.87 &    15.49 &  $    20.89\pm    3.16 $  &  $    1.31\pm    0.38 $ \\
$   3.00<|b | <   3.50$ &    3.23 &   931147 &   25 &    37.17 &    37.31 &  $    33.26\pm    7.46 $  &  $    2.21\pm    0.62 $ \\
$   3.50<|b | <   7.00$ &    4.45 &  2194599 &   24 &    36.75 &    41.51 &  $    32.60\pm    8.47 $  &  $    0.92\pm    0.27 $ \\
\\
$  -3.50<b <  -1.40$ &    2.69 &  1861880 &   54 &    35.68 &    30.22 &  $    31.33\pm    4.11 $  &  $    2.42\pm    0.47 $ \\
$   1.40<b <   3.50$ &    2.50 &  1512735 &   42 &    27.27 &    23.42 &  $    23.08\pm    3.61 $  &  $    1.94\pm    0.42 $ \\
\\
$  -6.00<\ell  <  -3.00$ &    2.40 &   361293 &   10 &    46.02 &    28.13 &  $    39.43\pm    8.89 $  &  $    2.90\pm    1.30 $ \\
$  -3.00<\ell  <   0.00$ &    2.42 &   660722 &   20 &    29.79 &    23.56 &  $    25.65\pm    5.27 $  &  $    2.32\pm    0.73 $ \\
$   0.00<\ell  <   3.00$ &    2.22 &   981742 &   31 &    27.38 &    21.75 &  $    23.77\pm    3.91 $  &  $    2.20\pm    0.56 $ \\
$   3.00<\ell  <   6.00$ &    2.53 &   316650 &    8 &    25.93 &    11.89 &  $    23.88\pm    4.20 $  &  $    1.65\pm    0.83 $ \\
\\
\end{tabular}
\caption{
The measured optical depth for various slices in Galactic latitude $b$ and Galactic longitude $\ell$.
}
\label{slicetable}
\end{center}
\end{table*}

\begin{figure}
 \centering
 \includegraphics[width=7.8cm]{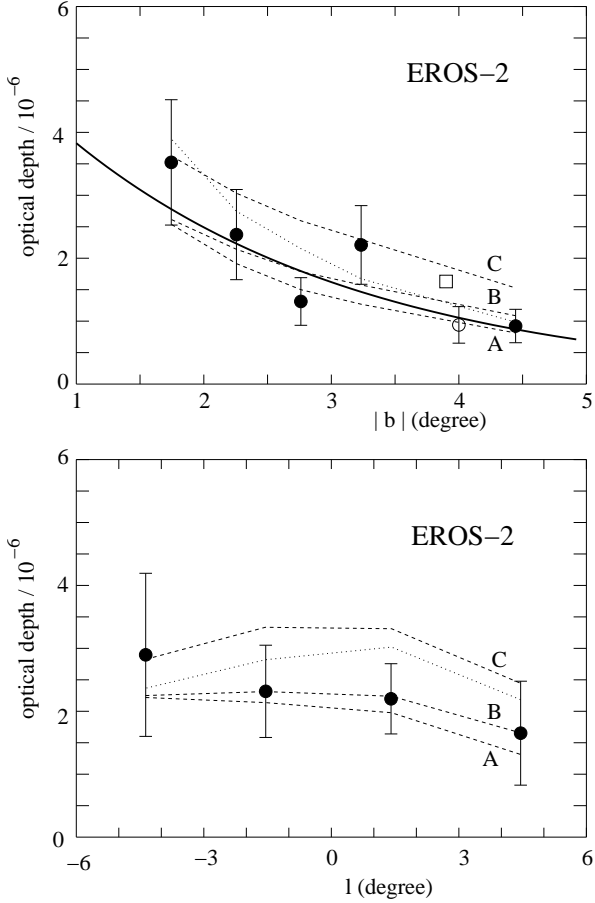}
 \caption{The top panel shows the EROS-2 measured 
optical depth as a function of Galactic latitude.
The bottom panel shows the optical depth as a function of
Galactic longitude for the latitude range $1.4\degr<|b|<3.0\degr$.
The filled circles are from this work while the open circle
in the upper panel is from the first EROS-2 analysis \citep{Afonso}.
In the upper panel, the solid line shows the fit (\ref{profoptfit}).
In both panels, the dotted lines show the prediction of the model of 
\citet{Bissantz} 
and the dashed lines, A, B and C, show the predictions
of three models used by  
\citet{EVA02} 
as described in 
Section \ref{conclusionsec}.
The open square in the upper panel 
is the prediction for the Baade Window  by 
\citet{HanGould03}.
}
  \label{profoptfig}
\end{figure}

Because of the rapid variation of the optical depth across
our fields, the above mean optical depth is of limited interest.
Table \ref{slicetable} 
and Figure \ref{profoptfig} shows the depth for various slices in 
Galactic latitude and longitude.
The latitude gradient is clearly seen.
Fitting the five points shown in Figure \ref{profoptfig}a, one finds
\begin{equation}
\tau/10^{-6}=(1.62\,\pm 0.23)\exp[-a(|b|-3\degr)] \;,
\label{profoptfit}
\end{equation}
with
\begin{equation}
a\;=\;(0.43\,\pm0.16)\,\deg^{-1} \;.
\end{equation}
At $b=-2.7\,\degr$ this corresponds to an optical depth gradient of
$(0.78\pm0.27)\times 10^{-6}\,\deg^{-1}$, in agreement with the less
precise values of MACHO \citep{machocgpop}
$(1.06\pm0.71)\times 10^{-6}\deg^{-1}$ and of OGLE-II \citep{sumiogle}
$(0.78\pm0.84)\times 10^{-6}\deg^{-1}$.
We see no significant difference between the northern and southern fields,
as expected.

The dependence on Galactic longitude is expected to be much weaker.
Additionally, nonuniform placement of the observing fields makes it
easy to mask a longitudinal dependence by the stronger latitude dependence.
Figure \ref{profoptfig}b shows the measured optical depth as a function
of longitude for EROS-2 fields in the latitude range $1.4\degr<|b|<3.0\degr$.
Also shown are predictions for our fields of various models.
No significant longitudinal dependence is seen.

\begin{figure}
 \centering
 \includegraphics[width=7.8cm]{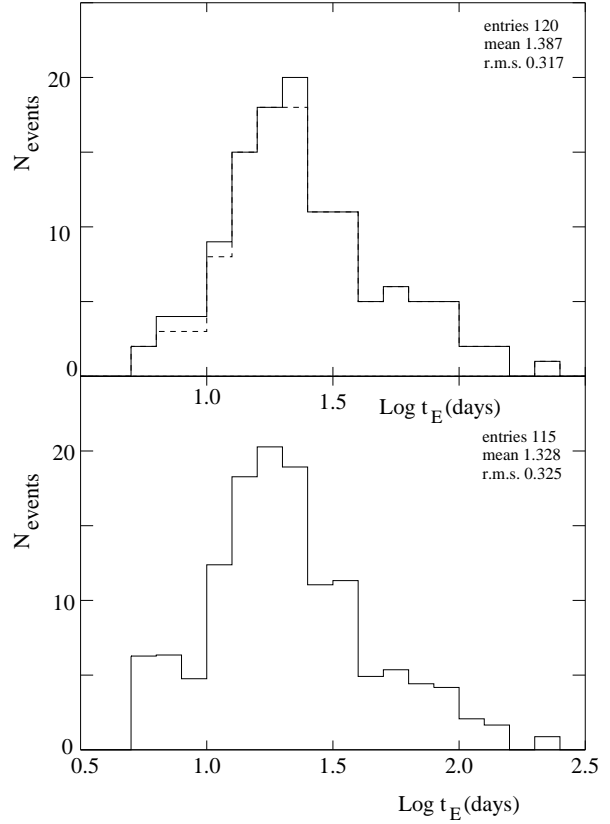}
 \caption{
The distribution of Einstein radius crossing times, $t_\e$.
The top panel shows the raw distribution for the 120 events
with $u_0<0.75$ with the dashed line corresponding to the 115 events
that show no strong blending.
The bottom panel shows the distribution corrected for 
the $t_\e$ dependence of the detection
efficiency.
}
  \label{tedistfig}
\end{figure}

\begin{figure}
 \centering
 \includegraphics[width=7.8cm]{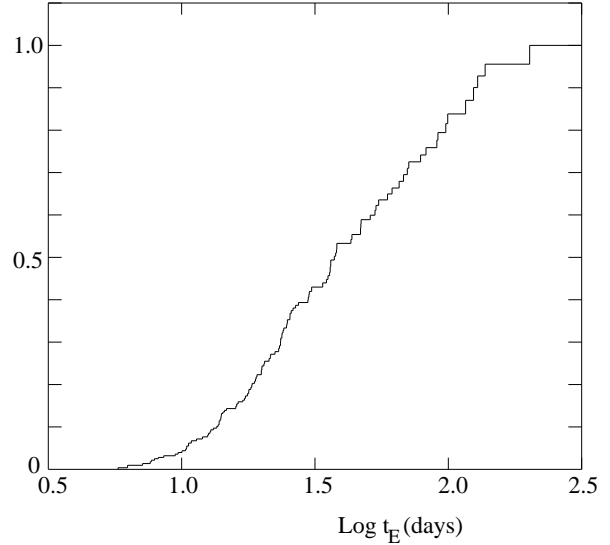}
 \caption{The normalized cumulative distribution of the
contribution of the 120 candidates to the observed
optical depth as a function of $t_\e$.
10\% of the optical depth comes from 
the 24 events with $t_\e < 13.6\,{\rm d}$ and 10\% from the
4 events with
$t_\e > 120\,{\rm d}$. 
}
  \label{teintfig}
\end{figure}

Figure \ref{tedistfig} shows the $t_\e$ distribution of the 120 events,
both
raw and efficiency corrected.  The mean $t_\e$ (Table \ref{slicetable})
is
\begin{equation}
\langle t_\e \rangle \;=\; (28.3 \,\pm\, 2.8)\,{\rm d} \;.
\label{meanteeq}
\end{equation}
We see no statistically significant latitude or longitude
dependence of $\langle t_\e \rangle$ (Table \ref{slicetable}).
The north-south difference,
\begin{equation}
\langle t_\e \rangle_{b>0} \,-\,\langle t_\e \rangle_{b<0} \;=\;
(8.2\,\pm\,5.5)\,\day \;,
\end{equation}
is not significant.
Figure \ref{teintfig} shows the cumulative $t_\e$ distribution
of the 120 events showing their contribution to the optical depth;
10\% of the optical depth comes from 
the 24 events with $t_\e < 13.6\,{\rm d}$ and 10\% from the
4 events with
$t_\e > 120\,{\rm d}$.

The 15\% error in the optical depth (\ref{profoptfit}) 
and 10\% error in $\langle t_\e \rangle$ (\ref{meanteeq}) are
mostly statistical, reflecting the number of events and the
distribution of $t_\e$.  We believe that it is unlikely that
any systematic errors are this large.

The most obvious systematic error would come from possible
contamination of the event sample with events not due to microlensing.
This is a serious problem for microlensing searches in the  Magellanic
Clouds.
We note however that  the Magellanic microlensing rate 
is at least a factor 10 lower than
in the Galactic Bulge. Furthermore,  
the identified Magellanic background events should not
be a problem here since they concern stars that are not
in the clump region (blue bumpers, Be stars) or background supernovae
that have light curves sufficiently different from microlensing
light curves to be easily identified for the high signal-to-noise ratio
events considered here.  At any rate,
there is no indication that our sample of 120 events is contaminated
with events not due to microlensing.  Among the indications
for lack of significant contamination are

\begin{itemize}

\item  The $\chi^2$ distribution (Figure \ref{c2fig}) indicates that the
function (\ref{noblendeq}) is a good description of the
light curves.

\item The $u_0$ distribution is nearly 
flat as expected (Figures \ref{u0distfig} and \ref{kolgofig}).

\item  The events show no sign of chromaticity (Figure \ref{chromfig}).

\item  The light curves are symmetric about the 
time of maximum amplification (Figure \ref{updnfig}). 

\item The optical depth  has the expected  dependence on 
the Galactic latitude
(Figure \ref{profoptfig}). 
A background of variable stars might be
expected to yield a 
measured optical depth that is latitude-independent.

\end{itemize}

A second systematic error could come from the
contamination of the source sample by stars 
not in the Galactic Bulge.
The number of source stars that are not in fact in the Galactic
Bulge can be estimated from the general luminosity function
and standard Galactic models.
Generally, one finds that less than  1\% of the stars in the
clump box (Figure \ref{cmdfig}) could be  foreground main sequence
stars. For negative latitudes, a contribution at the percent level
can also be expected from bright giants
in the Sagittarius
dwarf galaxy where stars are $\sim2.3\,{\rm mag}$ dimmer than
low-latitude stars in the Galactic Bulge \citep{alard1996}.

A third effect comes from the 
preferential selection of source stars on the near side of the Bulge
because the magnitude cut on source stars (Figure \ref{cmdfig})
favors stars on the near side.
Compared to a star at the Galactic center, stars on the near (far) side
of the bulge are shifted upward (downward) in the color-magnitude diagram.
Faint stars on the near
side are moved into the accepted magnitude range while bright stars move out.
Since there are more dim stars than bright stars, there is a net gain
of stars on the near side.  The opposite effect occurs on the far side
where there is a net loss of stars. 
We used a Monte-Carlo calculation to estimate 
the mean position of our source stars 
and found that it is  shifted towards us from the Galactic center by
about 5\% of the r.m.s. Bulge light of sight thickness.
Since stars on the near side 
have a lower than average optical depth for bulge-bulge lensing,
we can expect that the optical depth is underestimated
by of order 5\%.
This factor is not negligible compared to the statistical 
errors and any precise comparison with Galactic models will
require that models properly weight source stars by their 
position in the Bulge.

A fourth effect could be the 
misclassification of non-standard microlensing events (e.g. caustic events)
as variable stars.  Our detection criteria are sufficiently
liberal that several clear binary events were found, making
up about 8\% of the optical depth (Table \ref{scantable}).  The
ambiguous single excursion events would be only about 1\% of the
optical depth if they were included in the calculation.  It therefore
seems unlikely that there is a loss of events at the 10\% level.

A fifth systematic error could come from the
use of the  $t_\e$  fitted with (\ref{noblendeq}) 
for events due to non-standard lenses.
The problem is minor
for parallax events for which the modifications of (\ref{noblendeq})
are relatively small and the use of the fitted $t_\e$ makes little
difference.  On the other hand, caustic events have light
curves that resemble (\ref{noblendeq}) only in the wings and
the fitted $t_\e$ can be far from its true value, though for
the caustic events in this study, the differences are only 
of order 10\%.  
At any rate, the number of events of this category
is small, so we can expect
that the modification of the calculated optical depth should
be much less than 10\%.
This was the conclusion of the detailed study of 
\citet{Glicens}.

The sixth systematic effect is that due to
blending, the subject of the next section.

\section{Effect of blending on the measured optical depth}
\label{section:effect_blending}

Stellar blending complicates
the interpretation of microlensing events in crowded fields.
It is helpful to divide blending into two effects:

\begin{itemize}

\item Effect I: 
The photometry of bright (clump giant) stars is affected by the 
background of faint stars that are randomly placed with respect to the 
bright star.  
In the photometry, the sky background is assumed to be a smooth function so  
if the number of faint stars inside the seeing disk of the bright star 
is greater than the average number,
the baseline flux of the bright star will be overestimated and
the amplification due to  microlensing underestimated.
The opposite will occur if the number of faint stars is smaller than average.
We can therefore expect that on average amplifications are
correctly estimated.  However, as emphasized in the previous section,
the magnitude cut (Figure \ref{cmdfig}) used to select
source stars favors stars whose baseline fluxes are overestimated.
We therefore expect a slight underestimation of the amplification.

\item Effect II: 
The microlensing of a faint star within the seeing disk of 
a bright star will cause the reconstructed
flux of the bright star to vary in time, yielding an apparent microlensing
of the bright star.

\end{itemize}

By itself, effect II will clearly cause one to overestimate the optical
depth if (\ref{eqn:theo_opt}) is used since some of the events
will not be due to the $N_{\rm star}$ clump giants.  On the other hand,
we expect that  effect I goes in the opposite direction
since  amplifications of clump giants 
are on average underestimated.  Because of this,
the reconstructed $t_\e$ found using (\ref{noblendeq})
are systematically underestimated
while the reconstructed $u_0$ are overestimated.
In the formula (\ref{eqn:theo_opt}) for the optical depth, the
terms in the sum are 
therefore underestimated on average.  Furthermore, fewer events are
included in the sum because the overestimation of  $u_0$ causes
some events to migrate beyond $u_0(max)$.
Effect I therefore, by itself,
causes the optical depth to be underestimated.

To estimate the importance of blending we have performed
two types of tests with artificial stars placed on CCD images.
The first type uses totally synthetic images in which all stars are
artificial.  This was used to quantify the sum of effects I and II.  
The second type uses real EROS-2 images in which artificial clump
giants are added at random places on the images.  This was used
to quantify effect I.

The production and analysis of the totally synthetic images were
described in the first EROS-2 Galactic Bulge article \citep{Afonso}.
Series of CCD images were produced by placing on them  stars
taken randomly from 
the luminosity distribution observed for the Baade window 
in HST deep images \citep{HOLTZ98}.
The HST luminosity distribution extends to 9 magnitudes fainter
than clump giants and therefore allowed us to simulate the bright
stars used in the EROS-2 analysis as well as the fluctuating
background of faint stars.
Microlensing events were  assigned to a subsample of stars 
and the resulting flux change was taken into account when
placing charge on the CCD.
The sequence of CCD images were then reconstructed in the same
way as normal EROS-2 images and events were searched for.

The results of this study indicated that reconstructed amplifications are
indeed underestimated for clump giant stars leading to an
underestimation of the optical depth (effect I) but that this
is compensated to a precision of about 5\%
by events due to background stars (effect II).
For the analysis of 
\citet{Afonso},
about 18\% of the  recovered optical depth was due to these faint stars
compensating for the 15\% underestimation for those events due
to bright stars.  For the analysis presented here, the effects
are smaller because brighter stars are used and the amplification
threshold is higher (1.6 instead of 1.34).
We now find that only about 6\% of the recovered optical depth
is expected to be due to faint background stars.

Our second test used partially synthetic images on which
we placed artificial
clump giants on the real EROS-2 CCD images.  
As in the first test,
microlensing events were
simulated by giving the artificial giants time-dependent 
fluxes.  These so-called spiked images were then photometered by the normal
procedure and the events recovered and fitted with the microlensing
light curve.

\begin{figure}
 \centering
 \includegraphics[width=7.8cm]{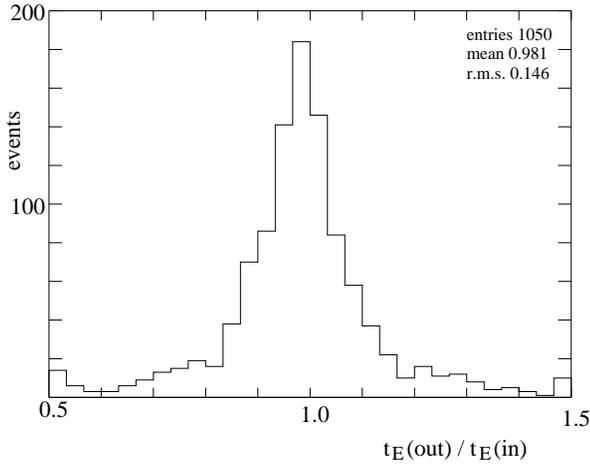}
 \caption{
The ratio of recovered and generated values of $t_\e$
for artificial clump giants placed on EROS-2 field 610.
}    
  \label{fig:mcbref}
\end{figure}

Figure \ref{fig:mcbref} shows the comparison between recovered and
simulated $t_\e$.  This parameter is  correctly
estimated on average 
with 85\% of events yielding a $t_\e$ within 20\%
of the generated $t_\e$.  The 15\% of events in the wings are
due to stars whose baseline flux was strongly overestimated or
underestimated because of blending.
The nearly symmetric shape of the distribution in Figure \ref{fig:mcbref}
means that a bias in the reconstruction of $t_\e$ results
mostly from the fact that
the magnitude cuts favor stars whose flux was overestimated.
The distribution 
in Figure \ref{fig:mcbref} 
and the associated correlation with reconstructed baseline flux
was thus combined with the observed magnitude distribution
to make a Monte-Carlo estimate of the optical depth bias.
The results confirm that effect II causes
$\sim5\%$ underestimation of the optical depth.

Since these images only give information on effect II,
we supplement it with a simple numerical calculation 
to estimate the contribution of faint stars in the seeing disk.
A microlensed faint star superimposed on
a bright star yields a light curve given by 
\begin{equation} 
F(t) \; = \; F_{\rm s}\left[ (1-f)
\,+\, f\,\frac{u^{2}+2}{u\sqrt{u^{2}+4}} \right]  \;,
\label{blendedcurve}
\end{equation}
where $f$ is the ratio between the faint star's baseline flux
and the total reconstructed baseline flux.
For $f\ll 1$, the reconstructed  amplification
is much less than the real amplification yielding
reconstructed $u_0$ and $t_\e$ respectively much greater than
and much less than the real  $u_0$ and $t_\e$.
In practice, this means that stars more than 4 magnitudes
fainter than clump giants yield no observable events
because their reconstructed $t_\e$ are less than $5\,{\rm d}$.
The HST luminosity function gives on average about 3 stars within
$2\,{\rm arcsec}$ of a clump giant and 
within 4 magnitudes of the clump magnitude.
Integrating the HST function, we can expect
that about 10\% of clump giant events will be due to faint background
stars.  Since the events are much shorter than real clump giant
events, they contribute only about 4\% to the measured optical depth,
not far from the above estimates.

Because of the approximate compensation,
we  make no blending correction to the optical depth.
Instead,   
because of  the difficulty in estimating the effect precisely, 
we choose to include a 5\% uncertainty in the optical depth,
i.e. an
uncertainty equal to the estimated size of each of the compensating
effects.
This uncertainty is, at any  rate, less than the statistical
uncertainty of the present measurements.  Future high statistics
measurements of the optical depth will certainly have to be much
more careful about blending, especially if faint source stars
are used.

Some strongly blended events can be found by fitting light curves
to a microlensing curve (\ref{blendedcurve}).
Since all events are blended to some extent, the number 
of detected blends is a sensitive function of the photometric
precision.
Five EROS-2 events showed a significant improvement for this curve
over that with the simple curve (\ref{noblendeq}).
The events are characterized by excess amplification in their wings
due to the fact that the real $t_\e$ is greater
than the $t_\e$ fitted with (\ref{noblendeq}).

\begin{figure}
 \centering
 \includegraphics[width=7.8cm]{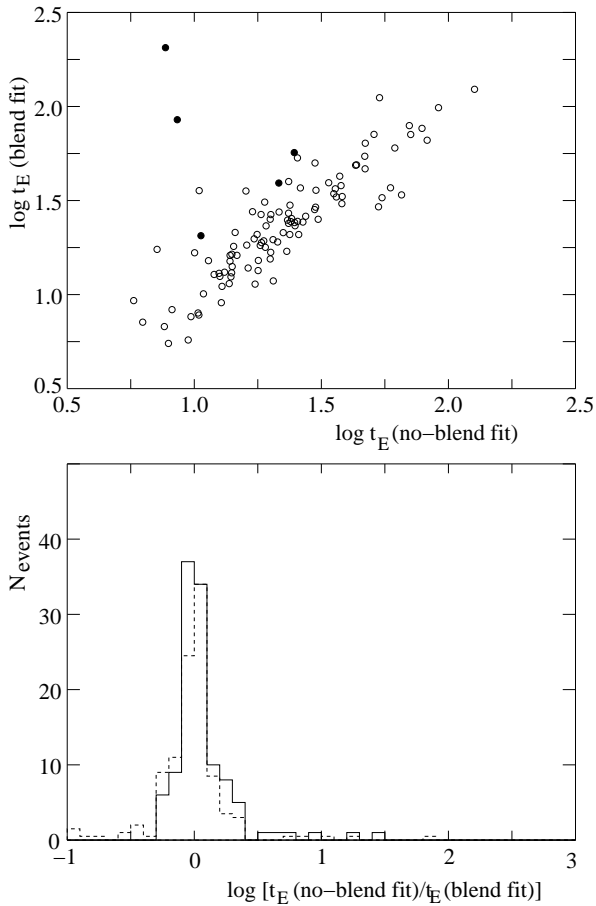}
 \caption{The fitted $t_\e$ assuming blending (equation \ref{blendedcurve}),
vs. the fitted value  
assuming no blending (equation \ref{noblendeq}).
The filled circles correspond to the five  events showing strong
blending.
The bottom panel shows the logarithm of the ratio of the two $t_\e$ for
simulated events (solid line) and observed events (dashed
line).
}    
  \label{fitblendfig}
\end{figure}

The absence of strong blending in most events is seen in
Figure \ref{fitblendfig}a
showing $t_\e ({\rm blend})$ {\it vs.}  $t_\e ({\rm no-blend})$.
While the cloud of points is very wide because of the degeneracies
in the fit, for the most part there is no systematic difference 
between the two $t_\e$, whereas 
$t_\e ({\rm blend}) >  t_\e ({\rm no-blend})$ for truly
blended events.
This is confirmed in
Figure \ref{fitblendfig}b, which shows the logarithm of the ratio of the
two $t_\e$'s both for the observed events and for the simulated
clump giants.  The observed distribution is very similar to
the Monte-Carlo distribution.

The absence of strong blending for our events is not inconsistent
with the results of the OGLE-II group who observed a large number
of strongly blended events on clump giants.  Most of the OGLE-II
blended events are very low amplification events, ($u_0({\rm no \; blend})>1$)
with only 4 of their 26 events with $u_0({\rm no\; blend})<0.75$ being 
characterized as blends. 
This proportion of blends is somewhat higher than our estimates,
a fact that
may be due to their superior photometric precision.
At any rate, the number is 
sufficiently small to be of minor importance for optical depth
determinations with $u_0<0.75$.

\section{Discussion and Conclusions}
\label{conclusionsec}

\begin{figure}
 \centering
 \includegraphics[width=7.8cm]{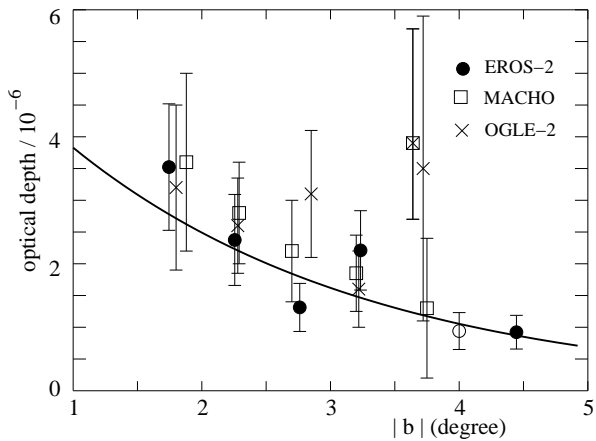}
 \caption{
Optical depth as a function of Galactic latitude.
The filled circles are this work while the open circle
is the first EROS-2 analysis \citep{Afonso}.
The solid line shows the fit (\ref{profoptfit}).
The measurements of the MACHO group are the
open squares \citep{machocgpop} and the crossed square of the early
analysis of 
\citet{ALC97}.
The crosses are from the OGLE-II group \citep{sumiogle}.
}
  \label{profoptallfig}
\end{figure}

The EROS-2 optical depth 
measurement presented here is in remarkably good agreement with
the other clump-giant measurements shown in Figure \ref{profoptallfig}, 
i.e. those of 
the MACHO group \citep{machocgpop} and OGLE-II group
\citep{sumiogle}.
Only the original low statistics MACHO measurement \citep{ALC97} 
is a bit too far  from our fitted optical depth (\ref{profoptfit}).

Figure \ref{profoptfig} shows that our optical depth measurements are in
reasonable agreement with some recent calculations based on models
of the Galactic Bar and Disk.
The dotted line in Figure \ref{profoptfig}  is the prediction of the 
non-parametric model of
\citet{Bissantz}.
The strong latitude dependence is correctly predicted as is 
the nearly flat longitude dependence.
The three dashed lines correspond to calculations of \citet{EVA02} based on
the three models of the inner Galaxy.  Curve A is based on the  model 
of \citet{Binmodel},  curve B on the model of  
of \citet{Dwek}, and curve C on the model of  
of \citet{Freu}. 
(For these three models, we show the predictions 
without corrections due to spiral structure;
these  would
give at most a 20\% increase to the calculated optical depth.)
All three models have a common total mass of $1.5\times10^{10}M_{\odot}$
within $2.5\,{\rm kpc}$ of the Galactic Center.  
The data in Figure \ref{profoptfig}
indicate that the ``swollen bar'' model of 
\citet{Freu} 
is not favored by the data, unless
the assumed mass is reduced by $\sim30\%$.

Finally, Figure \ref{profoptfig} shows the Baade window prediction of 
\citet{HanGould03}.  The calculated optical depth 
is within one standard deviation
of the curve given by (\ref{profoptfit}).

We make no attempt here to separate the Disk (\ref{optdepthdisk})
and Bulge (\ref{optdepthbulge}) contributions 
so our conclusion
is simply that the model of \citet{Bissantz}
or models A and B of \citet{EVA02} give a good estimate of
the total optical depth (\ref{optdepthint}).
We note that
these models place the bulk of the mass in normal stars and gas
so the agreement with our measurements indicate that there is no need 
for additional mass in a non-lensing form.  
 In models with cold dark matter, the dark
 component may make
 a significant contribution
 to the mass in the inner parts of the Milky Way.  For example, in
 the models of \citet{KZS}, the cold dark matter
 contribution to the rotation curve
 at $1\,{\rm kpc}$ from the Galactic Center is between 30\% and 50\%.
 Such a contribution would, by itself, lower the optical depth
 by roughly the same factor.
 Our optical depth measurement, with its 15\% uncertainty,
 therefore provides a useful constraint on such models.


Galactic models generally suppose a smooth distribution 
of lenses and therefore predict a smooth longitude and
latitude dependence of the optical depth.
The MACHO group \citep{machocgpop} observed 
9 events near $(\ell=2.9\degr,b=-2.9\degr)$ (their field 104), 
yielding an optical depth about 2 standard
deviations above the expectation based on their other fields.
As seen in Figure \ref{fig:cont_candidates},
we do not see an excess in this direction.
Of their 9 events, 4 events occurred after the beginning of EROS-2.
Of these 4 events, two are not in an Eros field and
two are in our field 611.   Both of these events were
found (in Table \ref{eventtable1}, MACHO events 104.20515.498
and 104.20640.8423).  No other EROS events were found in this region
in the three bulge seasons after the shutdown of MACHO.  
The fact that EROS-2
saw only two events for a solid angle and time period 
equivalent to that in which MACHO found 9 events
suggests that the MACHO excess is a statistical
fluctuation.

While our optical depth measurement is in good agreement with
other measurements and with Galactic models, 
agreement on  the $t_\e$ distribution
is less satisfactory.
Our value of $\langle t_\e \rangle$ is
in excellent agreement with the value,  $(28.1\,\pm4.3)\,{\rm d}$,
found by  OGLE-II \citep{sumiogle}.
The EROS-2 and OGLE-II values are, however, significantly
higher than that calculated using the 62 MACHO events
(Table 3 of \citep{machocgpop}):  
$\langle t_\e\rangle=(21.6\,\pm 3) \,{\rm d}$.
The discrepancy lies entirely in the large number of MACHO
events with $t_\e < 10\,{\rm d}$.
The number of MACHO events with ($t_\e<5$, $5<t_\e<10$, $t_\e>10\day$)
is $(6,14,42)$ whereas the corresponding numbers for EROS-2 
are $(0,10,110)$.  Part of the differences in event numbers is
due to the fact the the MACHO efficiency is relatively greater
than the EROS-2 efficiency
for $t_\e<10\,{\day}$. 
Using the 14 MACHO events with $5\,\day<t_\e<10\,\day$ and the relative
efficiencies of the two experiments, we estimate that EROS-2
should see $24\pm7$ events in this range compared to the 10
that are seen.  The very low EROS-2 efficiency for  $t_\e<5\,{\day}$
makes it unsurprising that we found no events in this range.

Whatever the source of the discrepancy,
these short events have relatively 
little effect on the measured optical
depths.  
The events with $t_\e<10\,{\rm d}$ constitute only about 12\% of
the MACHO optical depth and only 4\% of the EROS-2 optical depth.

The Galactic models are  capable of reproducing the 
$t_\e$ distribution by adopting an appropriate stellar mass function.
The model of \citet{woodmao} is in good agreement with our distribution
and with that of OGLE-II, both with $\langle t_\e \rangle\sim28\,\day$.  
The model of \citet{BI2004} is in good agreement
with the distribution reported by MACHO with 
$\langle t_\e \rangle\sim21\,\day$.  
According to \citet{woodmao}, the different $\langle t_\e\rangle$
predictions of the two models is mostly due to different adopted
mass functions leading to a mean lensing mass of $0.35M_\odot$ for
\citet{woodmao} and $0.11M_\odot$ for \citet{BI2004}.  Kinematical
differences between the two models may also have a non-negligible effect.

The present comparison with models is 
still primarily limited by the low number of
observed events.  Since the EROS-2 survey uses most of the 
fields in the Bulge region that are easily observed in optical bands,
a large increase in the number
of monitored clump giants would require infrared observations of highly  
obscured regions near the Galactic Center \citep{gouldkband}.
Of course the number of events can also be increased by using dim stars
but, in this case, reliable results would require a good understanding
of blending.

\bigskip

\begin{acknowledgements}
We thank  V.  Belokurov and N.W. Evans  for providing their microlensing
maps in computer-readable form and O. Gerhard for interesting discussions.
  We are 
  grateful to the technical staff of ESO, La Silla for the support
  given to the EROS-2 project. We thank J-F.~Lecointe and A.~Gomes for the
  assistance with the online computing and
the staff of the CC-IN2P3, especially the team in charge of
the HPSS storage system, for their help with the data management. 
AG was supported by grant AST-0452758 from the NSF and 
JA by the Danish Natural Science Research Council.
 
\end{acknowledgements}

\bibliographystyle{aa}


\clearpage
\onecolumn

\begin{longtable}{l l l l l l l l l l} \hline\hline
event & r.a & dec. & $b$ & $l$ & $t_\e$ & $u_0$ & $t_0$ & $\epsilon$ & note \\ 
\hline
\\
 607-1-4-  11480 &     270.2603 &     -29.0032 &   -2.942 &    1.698 &   12.780 &    0.689 &  1274.70 &   0.576 &     SC30-57488    \\
 607-3-3-   9841 &     270.2787 &     -29.3148 &   -3.110 &    1.441 &   17.354 &    0.691 &   716.60 &   0.577 &    M-113.19192.365  \\
 607-5-1-  14742 &     270.0417 &     -29.5630 &   -3.053 &    1.121 &   18.181 &    0.067 &  2113.40 &   0.713 &                   \\
 607-5-1-  29607 &     270.0288 &     -29.6350 &   -3.079 &    1.054 &   61.540 &    0.133 &   636.60 &   0.823 &    M-118.18797.1397 \\
 607-5-4-  20285 &     270.2924 &     -29.8155 &   -3.367 &    1.020 &   36.233 &    0.498 &  1315.50 &   0.810 &    M-118.19184.939  \\
 607-6-3-  29248 &     269.8908 &     -30.0002 &   -3.156 &    0.680 &   15.967 &    0.709 &  2137.00 &   0.560 &                   \\
 610-0-1-  17627 &     270.4365 &     -28.9678 &   -3.059 &    1.809 &   18.775 &    0.347 &  1775.80 &   0.671 &    SC30-636963    \\
 610-2-3-   4703 &     270.6194 &     -29.2331 &   -3.328 &    1.666 &   10.458 &    0.420 &  1988.40 &   0.497 &                   \\
 611-1-4-  24874 &     271.0400 &     -27.7431 &   -2.920 &    3.132 &   27.429 &    0.619 &   683.50 &   0.798 &    M-104.20515.498  \\
 611-3-4-  18476 &     271.1397 &     -28.1256 &   -3.184 &    2.849 &   13.708 &    0.699 &   998.40 &   0.706 &    M-104.20640.8423  \\
 612-7-4-   8538 &     271.8072 &     -32.0593 &   -5.593 &   -0.195 &   70.213 &    0.233 &  1019.10 &   0.867 &                   \\
 613-5-1-  23714 &     271.5512 &     -30.1737 &   -4.493 &    1.288 &   20.356 &    0.175 &  2144.10 &   0.722 &                   \\
 615-1-2-  22556 &     271.8863 &     -26.7402 &   -3.090 &    4.378 &   14.728 &    0.126 &  1599.30 &   0.650 &                   \\
 619-0-4-   7664 &     272.2137 &     -28.4529 &   -4.170 &    3.057 &    9.705 &    0.408 &   523.40 &   0.474 &    M-110.22455.842  \\
 619-5-4-  19755 &     272.7316 &     -29.0649 &   -4.862 &    2.769 &   16.290 &    0.507 &  1732.70 &   0.608 &                   \\
 619-7-4-  14204 &     272.7335 &     -29.4067 &   -5.026 &    2.478 &  116.099 &    0.264 &  1254.70 &   0.677 & P                 \\
 624-4-3-   3703 &     273.1066 &     -30.1842 &   -5.679 &    1.984 &   17.672 &    0.635 &  1414.90 &   0.588 &                   \\
 627-7-2-  14994 &     273.4693 &     -26.4997 &   -4.217 &    5.306 &   43.505 &    0.041 &  1958.20 &   0.692 &                   \\
   2-0-1-  13467 &     267.8171 &     -30.2087 &   -1.710 &   -0.441 &   59.171 &    0.519 &  2383.30 &   0.793 &                   \\
   2-0-4-  18520 &     268.1808 &     -30.4511 &   -2.103 &   -0.482 &   22.388 &    0.239 &  1707.10 &   0.682 &                   \\
   2-1-1-   5329 &     268.3954 &     -30.1464 &   -2.109 &   -0.125 &   26.760 &    0.506 &  2122.60 &   0.783 &                   \\
   2-3-3-  30438 &     268.4314 &     -30.6361 &   -2.383 &   -0.524 &   38.098 &    0.552 &   996.30 &   0.908 &                   \\
   2-5-1-  20157 &     268.2981 &     -30.9266 &   -2.432 &   -0.831 &   97.853 &    0.088 &  1726.80 &   0.863 & C                 \\
   3-4-3-   2992 &     268.1774 &     -29.4347 &   -1.585 &    0.381 &   65.395 &    0.476 &  2100.30 &   0.782 &                   \\
   3-4-3-  27217 &     268.2445 &     -29.5728 &   -1.705 &    0.294 &   14.034 &    0.129 &  1685.60 &   0.582 &    SC37-645044    \\
   3-5-2-  20099 &     268.3380 &     -29.6770 &   -1.828 &    0.248 &   14.104 &    0.645 &  1759.70 &   0.554 &     SC3-371229    \\
   3-5-3-  26926 &     268.6611 &     -29.5536 &   -2.008 &    0.501 &   23.762 &    0.437 &  1257.30 &   0.706 &     SC4-522952    \\
   3-6-4-  15890 &     268.2890 &     -30.0199 &   -1.965 &   -0.066 &   38.201 &    0.066 &   955.90 &   0.489 &      SC3-91382    \\
   4-3-2-   5389 &     269.2294 &     -30.2536 &   -2.786 &    0.164 &   16.954 &    0.527 &  1283.20 &   0.822 &                   \\
   4-3-3-   2508 &     269.4168 &     -30.0798 &   -2.839 &    0.397 &   13.941 &    0.233 &  1411.40 &   0.753 &                   \\
   4-4-4-   9143 &     268.9565 &     -30.6300 &   -2.770 &   -0.280 &   21.513 &    0.044 &  1719.10 &   0.797 & B                 \\
   4-5-2-   8464 &     269.0497 &     -30.6251 &   -2.837 &   -0.233 &   17.209 &    0.456 &  1985.60 &   0.744 &                   \\
   4-7-1-   7864 &     269.1321 &     -30.8306 &   -3.002 &   -0.370 &  137.482 &    0.728 &  2086.20 &   0.930 & P                 \\
   4-7-1-  20818 &     269.1327 &     -30.9075 &   -3.041 &   -0.435 &   14.483 &    0.614 &  2180.60 &   0.707 &                   \\
   4-7-2-  19218 &     269.2468 &     -31.0564 &   -3.200 &   -0.509 &   37.820 &    0.688 &  1833.10 &   0.912 &    SC22-414328    \\
   5-4-1-   8951 &     268.8382 &     -29.0702 &   -1.897 &    0.993 &   10.424 &    0.724 &  1093.70 &   0.537 &                   \\
   5-4-3-  22912 &     269.0256 &     -29.1386 &   -2.073 &    1.020 &   30.119 &    0.492 &  2063.00 &   0.825 &                   \\
   5-4-4-  11338 &     268.9376 &     -29.2330 &   -2.055 &    0.900 &    7.704 &    0.640 &  1271.10 &   0.411 & B                 \\
   5-5-2-  19037 &     269.2080 &     -29.2700 &   -2.277 &    0.992 &    7.912 &    0.000 &  1651.60 &   0.411 &                   \\
   6-1-3-  21678 &     270.0438 &     -30.3994 &   -3.468 &    0.411 &   13.550 &    0.060 &  1763.90 &   0.625 & C                 \\
   6-5-3-  18380 &     270.1206 &     -31.0448 &   -3.844 &   -0.102 &   23.286 &    0.208 &  2077.50 &   0.688 &                   \\
   8-3-1-   1345 &     270.1013 &     -27.7379 &   -2.195 &    2.709 &   78.728 &    0.726 &  1714.00 &   0.907 &                   \\
   8-4-2-  21125 &     269.7872 &     -28.4151 &   -2.290 &    1.986 &   34.932 &    0.358 &   948.70 &   0.801 & X  M-401.48408.649  \\
   8-4-2-  21623 &     269.6680 &     -28.3593 &   -2.171 &    1.979 &   43.148 &    0.124 &  2129.90 &   0.887 &                   \\
   8-7-3-  16227 &     270.2820 &     -28.5282 &   -2.724 &    2.114 &   19.994 &    0.469 &  1301.90 &   0.803 &    SC30-165305    \\
   9-7-4-   7143 &     271.0161 &     -31.5437 &   -4.755 &   -0.118 &   50.982 &    0.510 &  1450.80 &   0.914 &                   \\
  30-2-3-  19447 &     273.9211 &     -28.5940 &   -5.557 &    3.712 &   23.998 &    0.398 &  1698.10 &   0.758 &                   \\
  31-3-3-  35856 &     274.4120 &     -27.2411 &   -5.308 &    5.095 &   13.824 &    0.318 &  1628.90 &   0.503 &                   \\
  61-0-3-   8256 &     274.4475 &     -22.4102 &   -3.067 &    9.292 &   37.373 &    0.654 &  1743.70 &   0.781 &                   \\
  70-3-1-  18517 &     265.4297 &     -34.2263 &   -2.068 &   -4.914 &   90.589 &    0.181 &  1707.40 &   0.964 & C                 \\
  70-3-1-  18797 &     265.3838 &     -34.2285 &   -2.036 &   -4.936 &   91.418 &    0.207 &  2400.50 &   0.961 &                   \\
  70-3-4-  28566 &     265.5989 &     -34.4414 &   -2.300 &   -5.016 &   29.865 &    0.425 &  1602.40 &   0.871 &                   \\
  70-4-1-  17316 &     265.0553 &     -34.5743 &   -1.989 &   -5.374 &    7.140 &    0.567 &  1428.90 &   0.330 &                   \\
  70-5-2-  24247 &     265.3672 &     -34.7616 &   -2.306 &   -5.388 &   71.090 &    0.061 &  1842.60 &   0.890 &                   \\
  70-6-2-  31993 &     264.9205 &     -35.1415 &   -2.197 &   -5.908 &   46.872 &    0.246 &  1456.10 &   0.587 &                   \\
  71-3-1-  21364 &     266.2818 &     -33.3554 &   -2.214 &   -3.797 &   47.100 &    0.073 &   892.10 &   0.911 &                   \\
  71-7-1-  28518 &     266.4633 &     -34.0936 &   -2.727 &   -4.332 &   36.263 &    0.177 &  2024.70 &   0.906 & C                 \\
  72-1-1-  22000 &     266.3057 &     -34.4337 &   -2.792 &   -4.688 &   18.317 &    0.131 &  2343.10 &   0.722 &                   \\
  72-4-2-  12551 &     265.8097 &     -35.1914 &   -2.841 &   -5.544 &   21.568 &    0.461 &  1442.30 &   0.804 &                   \\
  73-3-4-   7001 &     267.3455 &     -33.4813 &   -3.040 &   -3.418 &   14.286 &    0.043 &   896.30 &   0.739 &                   \\
  74-5-3-  16791 &     267.4200 &     -35.0934 &   -3.917 &   -4.730 &   23.496 &    0.180 &  2177.00 &   0.793 &                   \\
  76-1-3-  12658 &     268.3605 &     -31.6522 &   -2.845 &   -1.417 &   82.468 &    0.676 &  1046.30 &   0.900 &                   \\
  77-0-1-  20761 &     267.5647 &     -33.1794 &   -3.042 &   -3.065 &   18.932 &    0.487 &   928.60 &   0.711 &                   \\
  77-0-3-   7242 &     267.7949 &     -33.0623 &   -3.149 &   -2.862 &  128.844 &    0.151 &  1628.20 &   0.883 & P                 \\
  77-1-4-  10694 &     268.3326 &     -33.2374 &   -3.625 &   -2.765 &   24.832 &    0.419 &  2005.60 &   0.853 &                   \\
  77-4-1-  13517 &     267.7165 &     -33.8108 &   -3.473 &   -3.525 &   23.493 &    0.144 &  1426.00 &   0.856 &                   \\
  78-4-1-   3363 &     268.7127 &     -31.8944 &   -3.225 &   -1.461 &    8.176 &    0.430 &  1387.10 &   0.492 &                   \\
  79-2-1-   8900 &     268.4667 &     -33.0348 &   -3.620 &   -2.534 &   29.763 &    0.344 &  1680.00 &   0.832 &                   \\
  80-3-2-   7178 &     269.8741 &     -32.2297 &   -4.246 &   -1.216 &   35.887 &    0.049 &  1047.00 &   0.876 &                   \\
  80-4-2-  10880 &     269.4209 &     -32.6063 &   -4.100 &   -1.740 &   10.348 &    0.733 &  2040.10 &   0.548 &                   \\
  81-3-2-  24224 &     269.7945 &     -33.7675 &   -4.946 &   -2.545 &   30.017 &    0.253 &  1341.00 &   0.887 &                   \\
  82-2-3-  24164 &     270.3955 &     -32.2755 &   -4.652 &   -1.018 &   13.832 &    0.285 &  2357.80 &   0.631 &                   \\
  83-1-1-   8206 &     270.5674 &     -33.2471 &   -5.253 &   -1.758 &   13.159 &    0.180 &  1315.10 &   0.614 &                   \\
  84-4-1-  15036 &     271.1088 &     -33.0547 &   -5.556 &   -1.351 &   19.032 &    0.000 &  2176.00 &   0.727 &                   \\
 101-1-1-  20136 &     263.7159 &     -29.3984 &    1.743 &   -1.622 &   10.623 &    0.094 &  2351.60 &   0.292 & B                 \\
 102-0-3-  17623 &     262.5237 &     -29.4664 &    2.577 &   -2.223 &   25.501 &    0.152 &  1322.90 &   0.593 &                   \\
 102-3-1-  13451 &     262.6484 &     -29.7646 &    2.322 &   -2.420 &   12.616 &    0.290 &  1093.50 &   0.554 &                   \\
 102-5-2-  15625 &     262.7204 &     -30.2758 &    1.990 &   -2.822 &   24.927 &    0.139 &  2197.60 &   0.587 &                   \\
 102-5-3-  29920 &     262.9454 &     -30.2184 &    1.858 &   -2.671 &   33.810 &    0.586 &  1348.90 &   0.642 &                   \\
 103-1-4-  21106 &     263.1299 &     -28.1925 &    2.829 &   -0.858 &   23.528 &    0.310 &  2380.90 &   0.630 &                   \\
 103-3-4-  27325 &     263.1901 &     -28.5500 &    2.590 &   -1.137 &   10.030 &    0.662 &  1777.20 &   0.463 &                   \\
 104-3-3-  29860 &     262.2052 &     -29.1929 &    2.959 &   -2.134 &   12.518 &    0.142 &  2064.20 &   0.572 &                   \\
 104-3-4-  27383 &     262.1975 &     -29.3197 &    2.895 &   -2.246 &   20.467 &    0.575 &  1984.30 &   0.651 &                   \\
 105-5-1-  14483 &     261.1343 &     -29.5197 &    3.553 &   -2.900 &    7.628 &    0.682 &  1636.30 &   0.334 &                   \\
 105-7-4-  12278 &     261.2801 &     -30.0158 &    3.171 &   -3.256 &   25.769 &    0.356 &  1604.50 &   0.739 &                   \\
 106-1-4-  11177 &     265.1363 &     -27.0417 &    1.947 &    1.043 &   30.726 &    0.324 &  1624.90 &   0.565 &                   \\
 106-4-1-   8809 &     264.7175 &     -27.5637 &    1.985 &    0.404 &    5.775 &    0.298 &  1701.50 &   0.291 &                   \\
 106-4-4-  24841 &     264.8552 &     -27.8152 &    1.748 &    0.251 &   24.738 &    0.453 &  1073.50 &   0.785 & B                 \\
 106-4-4-  31028 &     264.8404 &     -27.8541 &    1.739 &    0.211 &   17.878 &    0.535 &  1623.80 &   0.687 &                   \\
 106-5-1-  15482 &     265.0711 &     -27.6027 &    1.699 &    0.531 &   19.899 &    0.122 &  1293.10 &   0.610 &                   \\
 106-5-1-  26077 &     264.9651 &     -27.6713 &    1.742 &    0.424 &   11.974 &    0.593 &  2099.10 &   0.462 &                   \\
 106-5-4-  24760 &     265.1471 &     -27.8188 &    1.527 &    0.381 &   25.477 &    0.093 &  2524.00 &   0.679 &                   \\
 106-6-1-  13061 &     264.6275 &     -27.9500 &    1.847 &    0.032 &   36.059 &    0.283 &  1687.00 &   0.508 & C                 \\
 107-0-2-  21213 &     263.4739 &     -26.7017 &    3.382 &    0.578 &  124.365 &    0.025 &  1250.10 &   0.776 &                   \\
 107-1-2-  24707 &     263.7571 &     -26.7073 &    3.166 &    0.702 &   16.090 &    0.404 &  2080.40 &   0.543 &                   \\
 107-3-3-  31150 &     263.9854 &     -26.9224 &    2.878 &    0.621 &   19.904 &    0.591 &  1309.50 &   0.754 &                   \\
 107-4-1-   8762 &     263.3894 &     -27.1358 &    3.210 &    0.166 &   35.398 &    0.489 &  1754.90 &   0.758 &                   \\
 107-4-4-  23806 &     263.7192 &     -27.3873 &    2.827 &    0.100 &   53.574 &    0.650 &  1055.00 &   0.881 &                   \\
 107-5-2-  26345 &     263.8999 &     -27.3882 &    2.692 &    0.182 &   14.064 &    0.197 &  1710.70 &   0.576 &                   \\
 107-7-1-  22583 &     263.8056 &     -27.5884 &    2.654 &   -0.033 &    8.574 &    0.075 &  1635.90 &   0.415 & B                 \\
 107-7-2-    167 &     263.9081 &     -27.6208 &    2.561 &   -0.014 &    9.466 &    0.220 &  1616.40 &   0.455 &                   \\
 108-3-1-  12859 &     266.2558 &     -25.8209 &    1.736 &    2.604 &   12.864 &    0.499 &  1658.60 &   0.585 &                   \\
 108-3-1-  27549 &     266.2520 &     -25.9237 &    1.685 &    2.513 &   11.375 &    0.197 &   888.20 &   0.504 &                   \\
 108-4-1-  13544 &     265.9802 &     -26.1798 &    1.760 &    2.169 &   53.105 &    0.689 &  1322.70 &   0.854 &                   \\
 108-4-1-  23818 &     265.7825 &     -26.2489 &    1.874 &    2.019 &   99.504 &    0.672 &  2532.60 &   0.823 &                   \\
 109-4-3-  15519 &     265.2998 &     -25.5322 &    2.621 &    2.417 &   17.836 &    0.417 &  1365.00 &   0.686 &                   \\
 112-0-2-  12627 &     266.7715 &     -23.3296 &    2.629 &    4.995 &   21.280 &    0.615 &  1317.10 &   0.639 &                   \\
 112-2-2-  25953 &     266.8246 &     -23.7485 &    2.371 &    4.655 &   20.012 &    0.700 &  1648.60 &   0.765 &                   \\
 112-5-2-  24472 &     267.1992 &     -24.0774 &    1.907 &    4.541 &   23.119 &    0.676 &  1011.60 &   0.558 &                   \\
 112-6-1-   4305 &     266.7831 &     -24.1610 &    2.190 &    4.279 &   10.871 &    0.293 &  2473.80 &   0.363 &                   \\
 112-7-3-  14923 &     267.3655 &     -24.2420 &    1.693 &    4.474 &   54.797 &    0.612 &  1813.50 &   0.789 &                   \\
 113-1-2-  28239 &     266.3130 &     -23.4511 &    2.926 &    4.681 &   23.803 &    0.066 &  1967.50 &   0.768 &                   \\
 113-7-1-  23134 &     266.3956 &     -24.3175 &    2.411 &    3.967 &   26.167 &    0.294 &   863.60 &   0.817 &                   \\
 114-3-1-  11753 &     265.5659 &     -23.5507 &    3.458 &    4.254 &   18.387 &    0.076 &  2044.50 &   0.728 &                   \\
 114-5-1-  16265 &     265.5462 &     -23.9263 &    3.276 &    3.919 &   67.940 &    0.548 &  1664.30 &   0.810 & X                 \\
 117-6-3-  13250 &     268.0484 &     -22.5049 &    2.042 &    6.293 &    6.270 &    0.082 &  1643.30 &   0.203 &                   \\
 117-7-1-  12630 &     268.2245 &     -22.5131 &    1.898 &    6.366 &   24.160 &    0.100 &  1794.70 &   0.761 &                   \\
 124-1-1-  16687 &     259.9039 &     -27.8925 &    5.362 &   -2.066 &   19.173 &    0.180 &  1410.20 &   0.688 &                   \\
 124-7-2-  20832 &     259.9118 &     -29.1101 &    4.664 &   -3.106 &  201.882 &    0.311 &  2075.30 &   0.854 & P                 \\
 124-7-3-  21344 &     260.0548 &     -28.9748 &    4.638 &   -2.925 &   47.000 &    0.295 &   918.50 &   0.851 &                   \\
\\
\caption{The 120 events used for the measurement of 
the optical depth. The first column gives the
identification: field, CCD, quadrant, star number.
The time of maximum amplification, $t_0$, is shown in the eighth
column as (JD-2,450,000).
The efficiency, $\epsilon$, used in the optical depth calculation is
shown in the ninth column.  
The final column signals events
of special character: P (parallax), X (xallarap or binary lens 
without caustic), C (caustic binary lens),
B (strong blend).
Event numbers in the final column 
preceded by M refer to MACHO events \citep{machocgpop}
and those preceded by SC refer to OGLE-II events \cite{sumiogle}.
}
\label{eventtable1}
\end{longtable}

\end{document}